\documentclass[fleqn,usenatbib]{mnras}

\usepackage{newtxtext,newtxmath}

\usepackage[T1]{fontenc}

\DeclareRobustCommand{\VAN}[3]{#2}
\let\VANthebibliography\thebibliography
\def\thebibliography{\DeclareRobustCommand{\VAN}[3]{##3}\VANthebibliography}

\usepackage{graphicx}	
\usepackage{amsmath}	

\title[Gas Accretion]{Modelling Star Cluster Formation: Gas Accretion}

\author[Karam \& Sills]{
Jeremy Karam$^{1}$\thanks{E-mail: karamj2@mcmaster.ca}
and Alison Sills$^{1}$
\\
Department of Physics \& Astronomy, McMaster University, 1280 Main Street West, Hamilton ON, L8S 4M1, CANADA
}

\date{Accepted XXX. Received YYY; in original form ZZZ}

\pubyear{2022}

\begin{document}

\label{firstpage}
\pagerange{\pageref{firstpage}--\pageref{lastpage}}
\maketitle

\begin{abstract}

The formation of star clusters involves the growth of smaller, gas-rich subclusters through accretion of gas from the giant molecular cloud within which the subclusters are embedded. The two main accretion mechanisms responsible for this are accretion of gas from dense filaments, and from the ambient background of the cloud. We perform simulations of both of these accretion processes onto gas-rich star clusters using coupled smoothed particle hydrodynamics to model the gas, and N-body dynamics to model the stars. We find that, for both accretion processes, the accreting star cluster loses some of its original mass while gaining mass from either the ambient background or the dense filament. The amount of mass lost from both these processes is small compared to the total mass of the cluster. However, in the case of accretion from a background medium, the net effect can be a decrease in the total mass of the cluster if it is travelling fast enough through the ambient medium ($> 4$kms$^{-1}$). We find that the amount of mass lost from the cluster through filamentary accretion is independent of the density, width, or number of filaments funneling gas into the cluster and is always such that the mass of the cluster is constantly increasing with time. We compare our results to idealized prescriptions used to model star cluster formation in larger scale GMC simulations and find that such prescriptions act as an upper limit when describing the mass of the star cluster they represent.

\end{abstract}




\section{Introduction}

Star forming regions throughout galaxies are found to comprise of dense molecular gas which can collapse to form stars. These stars form in clustered groups that can contain many stars gravitationally bound to one another (called star clusters). Such clusters are embedded inside these giant molecular clouds (GMCs) in their early stages of evolution (\citealt{ladalada}). This embedded phase of the star clusters life tends to last only a few Myr after which, the cluster is able to remove surrounding gas through stellar feedback effects (\citealt{li2019}, \citealt{pelupessy2013}). Interactions between the embedded star cluster and the surrounding molecular gas in this time can have an effect on the overall evolution of the star cluster. 

Because of turbulence, the distribution of the gas surrounding star clusters in GMCs can be very complicated. For example, observations performed by the Herschel telescope find that such star forming clouds are filled with filamentary structure (see \citealt{andre} and references therein). Furthermore, these filaments do not exist on their own, but are often found in hub systems, which contain many groups of filaments, that can surround young clusters or young stellar objects (YSOs) (e.g \citealt{SDC}, \citealt{fukui_papillon}, \citealt{kumar}, \citealt{bhadari}, \citealt{30_dor}). As these filaments evolve, they develop dense cores (e.g. \citealt{mens}, \citealt{arz2011}) within which stars and star clusters can form. The remaining filamentary gas can then be accreted onto the cluster (e.g \citealt{kirk}).

The sizes of these star clusters can vary, but numerical simulations of GMC evolution have shown that smaller star clusters (subclusters) can evolve into more massive ones with. Two important mechanisms for this growth are subcluster mergers, and accretion of the background gas in which the cluster is embedded (\citealt{Howard2018} hereafter H18). We focus on the latter in this work and point the reader to \citet{karam} (hereafter Paper I) for a study of subcluster mergers.

Simulating star formation inside GMCs is computationally challenging because of the small timesteps required to properly model the high density gas. Therefore, in order to model star cluster formation and evolution alongside the rest of the GMC, a sink particle prescription is often used. A commonly used prescription is that described in  \citet{sink}, which was used in H18. This approach models a subcluster of gas and stars as a sink with parameters that describe its overall behaviour (i.e. the total mass, position, and velocity) rather than resolving the cluster as a collection of individual stars and gas. Though sink particles drastically reduce computation time, a caveat to their use is that the internal evolution of the stars and gas that make up the star cluster is unknown. For example, in Paper I, we found that the sink particle prescription does not provide the full picture regarding subcluster mergers present in H18. We found that the merger of two clusters unbinds a fraction of the stars and gas from both clusters involved. Because sink particles do not account for this detail, we concluded that they could be overestimating the total mass of a system which is the result of a cluster merger (\citealt{karam}).

While simulations have been performed to study gas accretion processes onto clusters in more detail, they come with their own set of limitations. For example, simulations performed by \citet{naiman} have considered the star cluster as an analytic potential allowed to move through an initially uniform distribution of background gas. This method is similar to a sink particle approach in that it does not allow one to learn how the individual components of the star cluster (stars and gas) react to gas accretion. Others such as \citet{kaaz} have only considered clusters with up to $N = 32$ equal mass stars to understand accretion onto individual stellar members. This is not representative of true star clusters which can contain many more stars of varying mass.

In order to explore the effects of gas accretion on young star clusters in more detail, we model clusters at a higher resolution containing both stars and gas, and allow these components to gravitationally interact with one another. We then allow our cluster to interact with surrounding gas that would be present in a GMC. We consider accretion from ambient gas as the cluster moves through the GMC and from filaments that funnel gas towards the cluster. We analyze the cluster as it reacts to these processes to better understand how both the stellar and gas components evolve due to gas accretion inside GMCs and to use this understanding to inform large scale GMC simulations which use sink particle prescriptions to model star clusters.

In section \ref{sec:methods} we discuss the initial conditions used for both our stream and filament simulations. In section \ref{sec:cluster_evol}, we look at the response of the cluster to accretion from a ambient background media of different densities, and accretion from one, two and three filaments. Finally, in section \ref{sec:discussion}, we compare our results to observations to further understand star formation, and discuss how they can inform current implementations of the sink particle prescription.

\begin{table}
    \centering
    \begin{tabular}{c|c|c|c}
         Cluster Name & M$_\mathrm{star}$ [10$^4$M$_\odot$] & M$_\mathrm{gas}$ [10$^4$M$_\odot$] & r$_\mathrm{hm}$ [pc]\\
         \hline
         C1 & 0.06 & 0.3 & 0.4 \\
         C2 & 0.3 & 0.3 & 0.6\\
         \hline
         \hspace{1pt}
    \end{tabular}
    
    \caption{Parameters of the central clusters used in this work. Column 1: the name of the cluster, column 2: the total stellar mass of the cluster, column 3: the total gas mass of the cluster, column 4: the half mass radius of the cluster. Any simulation name from table \ref{tab:fil_sims} that does not have \_C2 after it, is using C1 as the central cluster.}
    \label{tab:clusts}
\end{table}

\begin{table*}
    \centering
    
    \begin{tabular}{c|c|c|c|c}
         Run Name & Ambient Background Density [M$_\odot$pc$^{-3}$] & Background Medium Velocity [kms$^{-1}$]\\
         \hline
         CtS005 & 0.05 & 4\\
         CtS05 & 0.5 & 4\\
         CtS1 & 1 & 4\\
         CtS005\_6v & 0.05 & 6\\
         CtS05\_6v & 0.5 & 6\\
         CtS1\_6v & 1 & 6\\
         CtS005\_10v & 0.05 & 10\\
         CtS05\_10v & 0.5 & 10\\
         CtS1\_10v & 1 & 10\\
         \hline 
         \hspace{1pt}
         \end{tabular}
         
         \caption{Parameters for our simulations of a star cluster moving through an ambient background medium. Column 1: the name of the simulation, column 2: the density of the ambient background medium (1M$_\odot$pc$^{-3}$ = 7$\times$10$^{-23}$gcm$^{-3}$ $\approx$ 17cm$^{-3}$), column 3: the velocity of the ambient background medium. All of these simulations use C1 as the central cluster (see table \ref{tab:clusts}).}
         \label{tab:CtS_sims}
\end{table*}
\begin{table*}
         \begin{tabular}{c|c|c|c|c}
          Run Name & Ambient Background Density [M$_\odot$pc$^{-3}$] & Filament Density [M$_\odot$pc$^{-3}$] & Filament Width [pc] & Number of Filaments \\
          \hline
         FwC1250 & - & 1250 & 0.3 & 1\\
         FwC850 & - & 850 & 0.3 & 1\\
         FwC600 & - & 600 & 0.3 & 1\\
         FwC310 & - & 310 & 0.6 & 1 \\
         2FwC1250 & - & (1250,1250) & 0.3 & 2 \\
         2FwC850 & - & (850,850) & 0.3 & 2 \\
         2FwC600 & - & (600,600) & 0.3 & 2 \\
         2FwC310 & - & (310,310) & 0.6 & 2 \\
         
         FCS1250\_1 & 1 & 1250 & 0.3 & 1 \\
         FCS1250\_05 & 0.5 & 1250 & 0.3 & 1\\
         2FCS1250\_1 & 1 & (1250,1250) & 0.3 & 2 \\
         2FCS1250\_05 & 0.5 & (1250,1250) & 0.3 & 2\\
         
         FwC1250\_C2 & - & 1250 & 0.3 & 1\\
         FwC850\_C2 & - & 850 & 0.3 & 1\\
         FwC600\_C2 & - & 600 & 0.3 & 1\\
         FwC310\_C2 & - & 310 & 0.6 & 1 \\
         2FwC1250\_C2 & - & (1250,1250) & 0.3 & 2 \\
         2FwC850\_C2 & - & (850,850) & 0.3 & 2 \\
         2FwC600\_C2 & - & (600,600) & 0.3 & 2 \\
         2FwC310\_C2 & - & (310,310) & 0.6 & 2 \\
         SDC & - & (850,600,600) & 0.3 & 3 
    \end{tabular}
    \caption{Parameters for our simulations of accretion from filaments onto a central star cluster. Column 1: the name of the simulation, column2: the density of the ambient medium, column 3: the density of the filament(s), column 4: the filament width, column 5: the number of filaments present in the simulation.}
    \label{tab:fil_sims}
\end{table*}

\section{Methods}
\label{sec:methods}
In the following sections, we discuss the computational methods used in our simulations. We also discuss the initial conditions of our two different simulation setups: a star cluster accreting gas from movement through a background medium, and a star cluster accreting gas from a dense filament.

\begin{figure}
    \centering
    \includegraphics[scale=0.24]{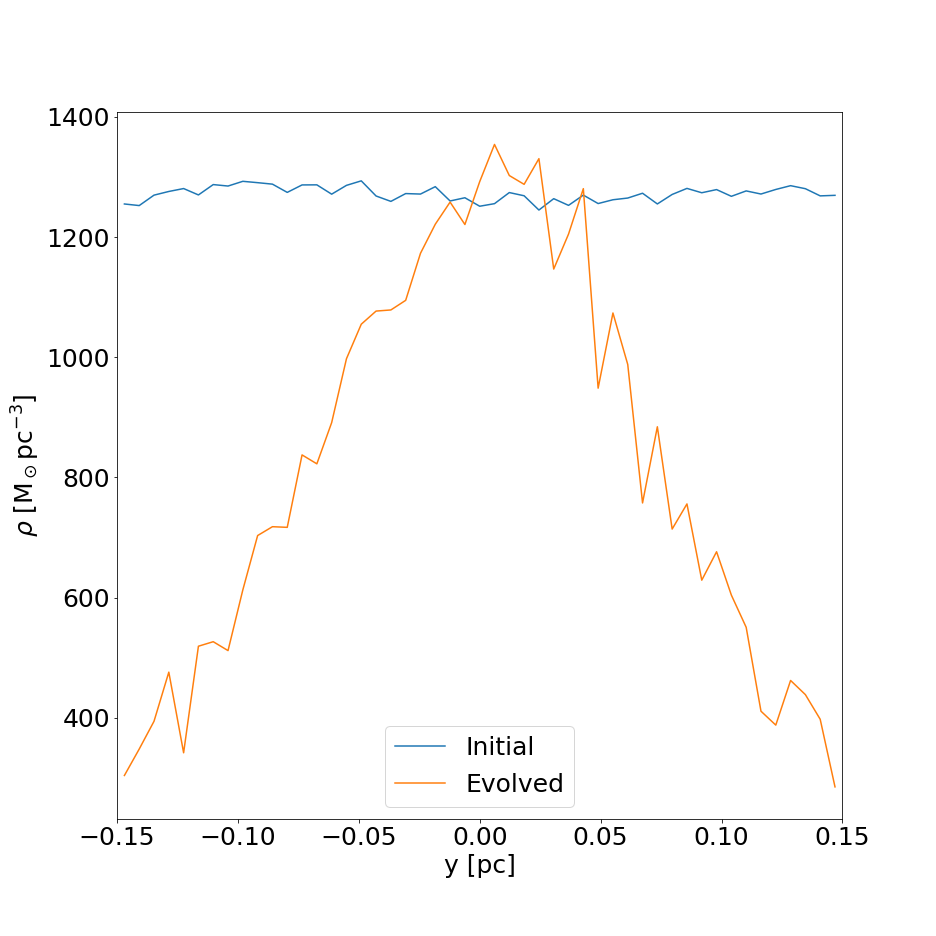}
    \caption{An example of filament density as a function of position across the y-axis of filament. We set up the filament with a constant density of $\rho=1250$M$_\odot$pc$^{-3}$ (blue line) and let it relax to a density profile given by the orange line after interaction with a point mass.}
    \label{fig:fil_dens_ex}
\end{figure}

\subsection{Numerical Methods}

Our simulations are performed using the Astrophysical Multipurpose Software Environment (AMUSE) (\citealt{zwart2009}, \citealt{pelupessy2013}) which contains codes that evolve the equations of gravity and hydrodynamics. AMUSE also allows for communication between these codes. This allows us to simulate the dynamics of the stars and hydrodynamics of the gas simultaneously in the cluster and focus on how they interact with one another. 

We use \texttt{hermite0} (\citealt{hermite}) for our N-Body code and for our smoothed particle hydrodynamics (SPH) code we use \texttt{GADGET-2} (\citealt{gadget2}). For the communication scheme, we use \texttt{BRIDGE} (\citealt{bridge}) with \texttt{BHTree} (written by Jun Makino based on \citealt{tree}) as the connecting algorithm. We describe our numerical scheme in more detail in Paper I.

We set up our star clusters using Plummer (\citealt{plummer1911}) spheres of stars and gas that we numerically relax. We assign the masses to the stars by sampling a Kroupa (\citealt{kroupa2001}) IMF from 0.15M$_\odot$ to 100M$_\odot$. For the gas, we give our SPH gas particles a mass of $M_{SPH} = 0.06$M$_\odot$ and set the gas temperature to $T = 10$K. We sample the velocities of the stars and gas using the method outlined in \citet{aarseth} and scale them to ensure that our clusters are initially in virial equilibrium (2$K_{s,g}/|P_{s,g}| = 1$ where $K_{s,g}$ is the kinetic energy of the stars or gas, and $P_{s,g}$ is the potential energy of the stars or gas). 

We ensure that the size of our cluster is such that the density at its half mass radius is consistent with young massive cluster observations as shown in \citet{zwart} ($\rho_{hm}$ $\approx$ 10$^3$-10$^4$M$_\odot$pc$^{-3}$). We show the parameters that describe the clusters used in this work in table \ref{tab:clusts}. For more information regarding our set up of the star cluster, see Paper I.

\subsection{Cluster Moving Through a Uniform Background}
\label{sec:CtS_methods}

We begin with describing our simulations of the motion of a cluster through an ambient medium of background gas. We place the cluster at the centre of the simulation box and set up a cube of constant density gas distributed in a glass configuration (\citealt{white_glass}). We give the background gas a velocity towards the cluster, thus placing our reference frame on the cluster as it moves through an ambient medium. We choose this velocity to be $v_{\mathrm{inj}} = 4$kms$^{-1}$ which is consistent with velocity dispersion values used in GMC evolution simulations (e.g. \citealt{starforge1}). We also study the effects of increasing this velocity to 6 and 10 kms$^{-1}$, the former being the average collision velocity of subcluster mergers in H18 and the latter being the velocity limit beyond which we found 
that cluster collisions may not result in a single merged cluster (see Paper I). As the background gas moves through the simulation box, we continuously inject material of the same density on the left side of the box so that we can study the cluster for longer times.

The size of the ambient background gas (and the simulation box) is 15x15x15pc. Because the cluster has a radius of $\approx 2$pc, this box size is sufficient to include all the original cluster material. Any gas or star particle that moves outside the boundaries of our simulation box is removed from the simulation. To study the response of our isolated cluster in varying environments throughout a GMC, we consider three different density values for our background gas: $\rho = 0.05, 0.5$ and $1$M$_\odot$pc$^{-3}$. This covers the low density regime used in previous works (e.g. \citealt{calura}) and the average densities observed in GMC catalogues from \citet{rosolowsky}. We show the parameters for all of our simulations of a cluster moving through an ambient background medium in table \ref{tab:CtS_sims}. Higher density regions throughout the GMC will likely be in the form of filaments which we cover in the next subsection.

\begin{figure*}
    \centering
    \includegraphics[scale=0.3]{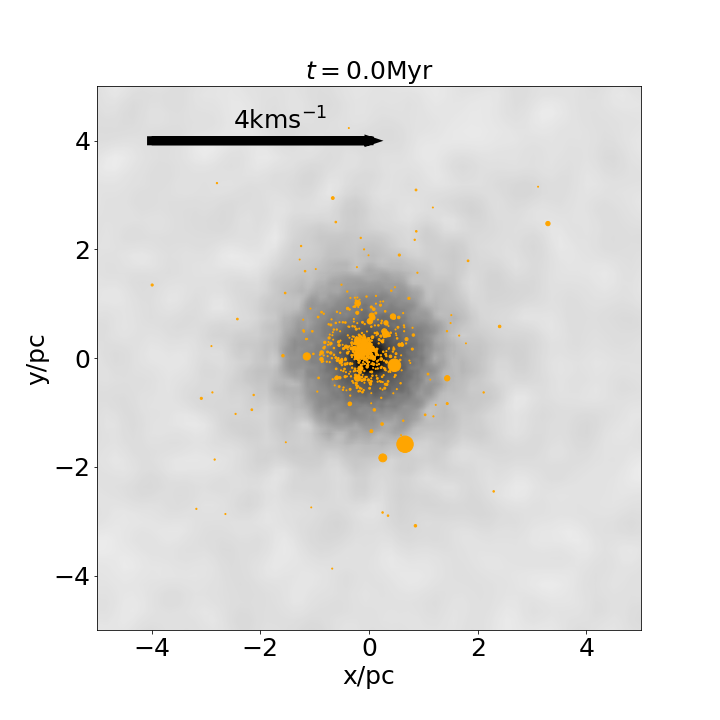}
    \includegraphics[scale=0.3]{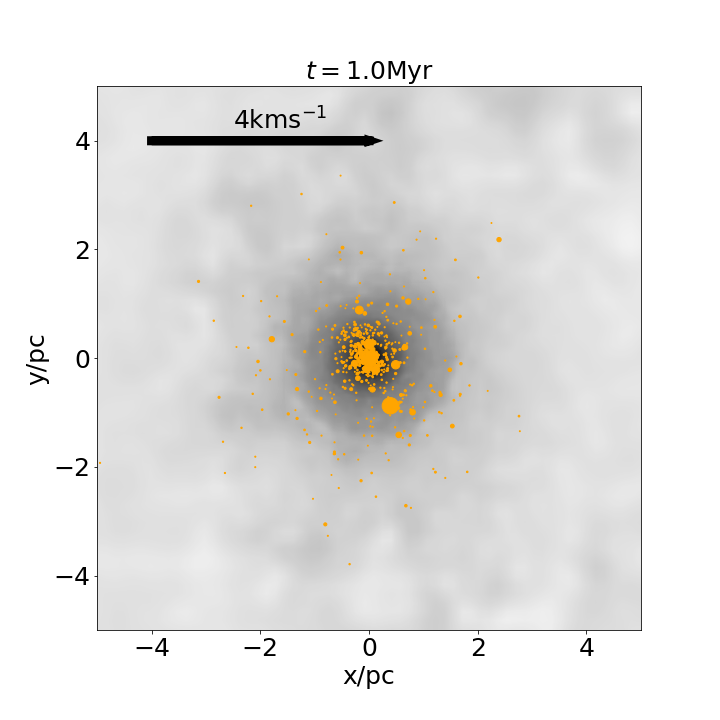}
    \includegraphics[scale=0.3]{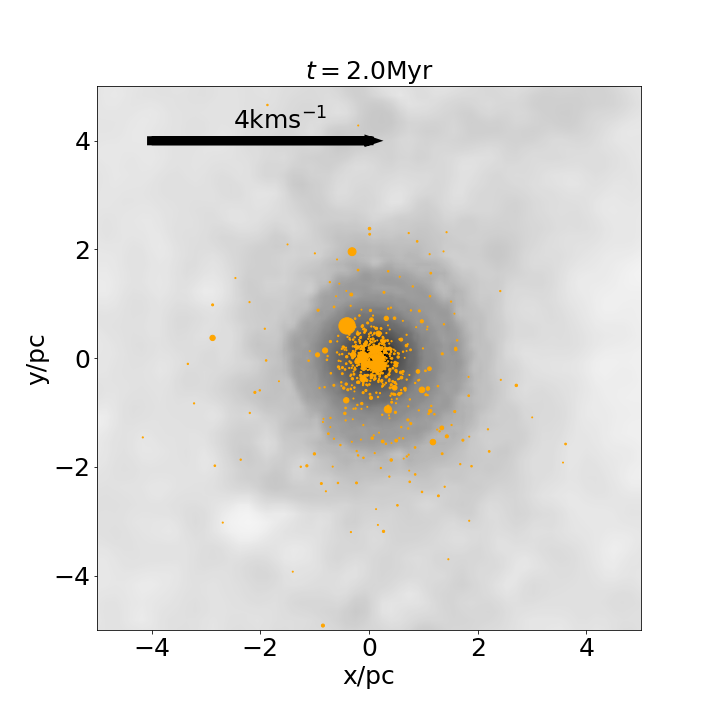}
    \includegraphics[scale=0.3]{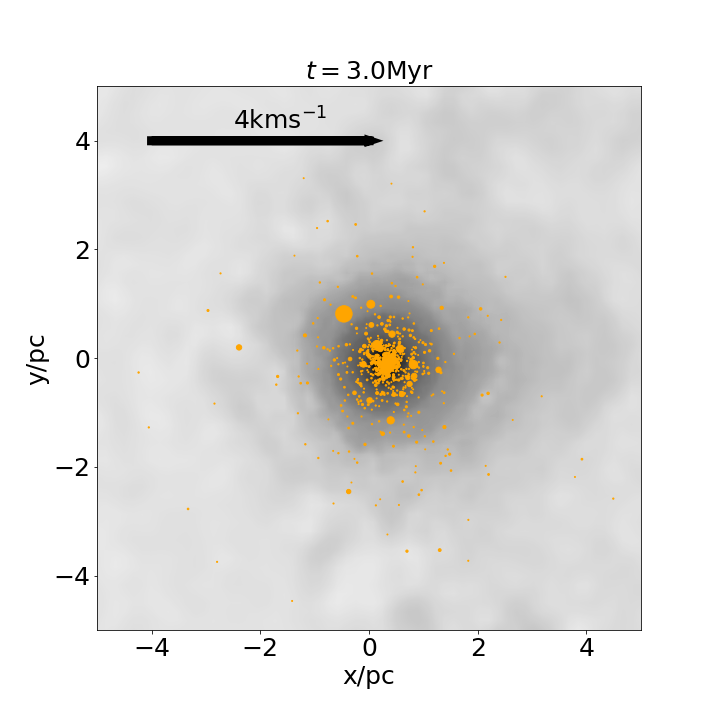}
    \caption{Snapshots of the stars and gas from ambient background gas accretion run CtS1. The orange circles represent the stars from the cluster with their size scaled to the mass of the star. Gas is shown in greyscale with darker regions showing higher density gas (minimum corresponds to $\rho=0.1$M$_\odot$pc$^{-3}$ and maximum corresponds to $\rho=10^4$M$_\odot$pc$^{-3}$). The black arrow in the top left shows the velocity of the ambient background gas.}
    \label{fig:CtS_snap}
\end{figure*}

\subsection{Filaments}
\label{sec:FwC}
The equilibrium state of a filamentary (thin and elongated) gas distribution is heavily dependant on the line mass ($M_{\mathrm{line}}$) of that distribution, defined as the mass per unit distance across the distribution (\citealt{inutsuka_1}, \citealt{inutsuka_2}). \citet{inutsuka_1} show that filaments whose $M_{\mathrm{line}}$ exceeds a critical value given by $M_{\mathrm{line}}^{\mathrm{c}} = 2c_s^2/G$ where $c_s$ is the thermal sound speed can collapse into dense, spherical cores. As well, \citet{fischera} show that filaments require sufficient external pressure to survive if their $M_{\mathrm{line}} < M_{\mathrm{line}}^{\mathrm{c}}$. We define a ratio of the line mass of a filament to the critical line mass as $f_{\mathrm{fil}} = M_{\mathrm{line}}/M_{\mathrm{line}}^c$. With this definition, $f_{\mathrm{fil}} > 1$ corresponds to filaments which will collapse under their own self-gravity, and $f_{\mathrm{fil}} < 1$ corresponds to filaments which require external pressure to remain stable.

We initially set up our filament as a thin, elongated distribution of SPH particles with uniform density $\rho_0$ and a width $w$. To give our filament a density distribution that more accurately resembles observations by \citet{arz2011} and theoretical models of cylindrical equilibrium states presented in \citet{ostriker}, we give our initial distribution a velocity ($v_{\mathrm{fil}}$) towards a point mass and let it evolve. An example of the change in our filament density profile can be seen in figure \ref{fig:fil_dens_ex} for a filament whose long axis is situated along the x-axis. In this example, we use the distribution given by the orange line as our filament that will accrete onto our central cluster. We take this filament and place it along the $y=0$ axis with the cluster situated at the center of our simulation box. As the filament travels towards the cluster, we continue to inject material from the edge of the filament to ensure that matter is continuously flowing.

We select filament densities and widths that are consistent with observations of star forming regions within the Milky Way, namely, the SDC13 (\citealt{SDC}), and IC5146 (\citealt{arz2011}) regions. These values have a direct effect on the line mass of our filaments because $M_{\mathrm{line}} = \rho_0\pi w^2$. Therefore, we distinguish between our filaments using $f_{\mathrm{fil}}$ throughout the rest of this paper. We give our filaments a velocity of $v_{\mathrm{fil}} = 2$kms$^{-1}$ which is consistent with simulations preformed by \citet{gomez_fil_vel}. We show the parameters for these simulations in table \ref{tab:fil_sims}.

Observations of filamentary regions give us a wide array of possible filament configurations that often include accretion onto a cluster from more than one filament. We therefore simulate this accretion from 1, 2 and 3 of our filaments with the orientation of our 3 filament simulation matching observations of SDC13 (\citealt{SDC}, \citealt{SDC_2}).

\section{Cluster Evolution}
\label{sec:cluster_evol}
In this section, we first discuss how our cluster is affected by movement through diffuse background gas. We then focus on how filamentary gas accretion impacts the cluster.

\subsection{Ambient Background Gas}
\label{sec:CtS}

We begin with a discussion of the evolution of our star cluster as it moves through ambient background gas with different velocities and different ambient background gas densities. These correspond to the simulations from table \ref{tab:CtS_sims}. We show snapshots of one of these simulations in figure \ref{fig:CtS_snap}. Here, our ambient background gas has a density of $\rho_{BG}=1$M$_\odot$pc$^{-3}$ and is travelling with positive velocity of 4kms$^{-1}$ along the x direction towards the cluster initiated at the centre of the simulation box. The cluster has a stellar mass of $M_s \approx 0.06 \times 10^4$M$_\odot$ and the gas mass is $M_g \approx 0.3 \times 10^4$M$_\odot$.

We find that the stellar component of the cluster is mostly unaffected by this evolutionary process regardless of ambient background gas density, or velocity. The gas is affected however. As the diffuse background gas pushes past the cluster, it starts to interact with the gas component of the cluster. At later times, we can start to see some of the gas being pushed away from the cluster (this can be seen as the dense gas directly to the right of the cluster in the bottom right snapshot of figure \ref{fig:CtS_snap}). This modifies the spherical symmetry that was present in the gas component of the cluster. Along with this loss of gas, the potential from the cluster leads to accretion of background gas implying that the total mass of the cluster is changing constantly throughout the simulation.

\begin{figure}
    \centering
    \includegraphics[scale=0.25]{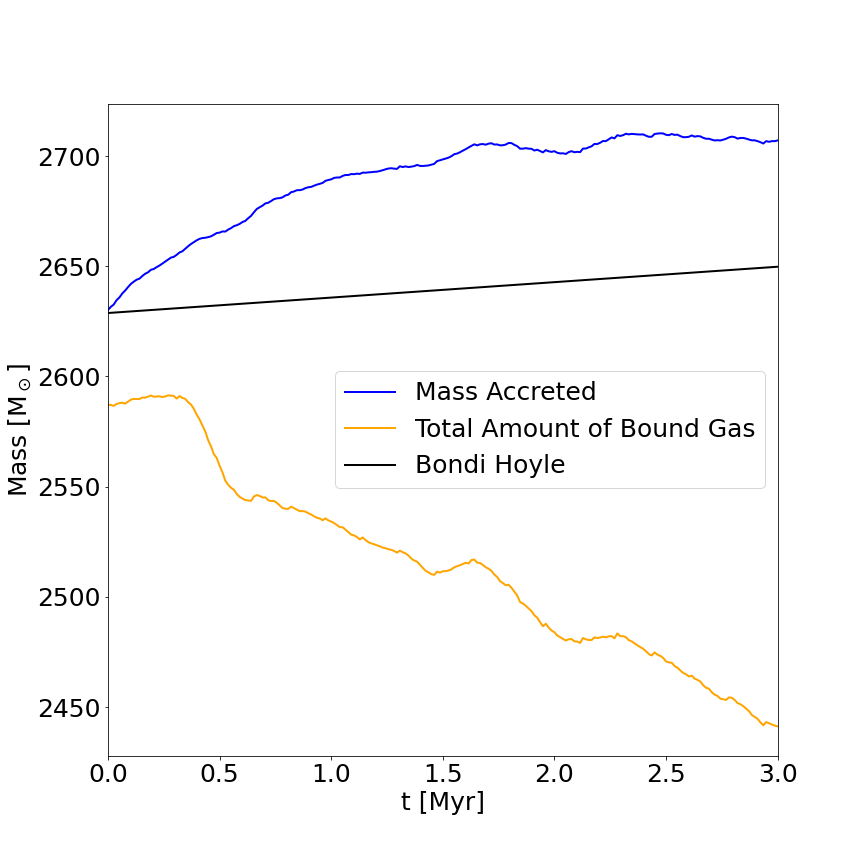}
    \caption{Total gas mass calculated assuming only accretion (blue line), using Bondi-Hoyle accretion formalism (black line), and using total bound mass of cluster (orange line) of a cluster moving through an ambient medium with a density of $\rho_{\mathrm{BG}} = 0.5$M$_\odot$pc$^{-3}$ and at a velocity of 6kms$^{-1}$.}
    \label{fig:stream_mass2}
\end{figure}

\begin{figure*}
    \centering
    \includegraphics[scale=0.45]{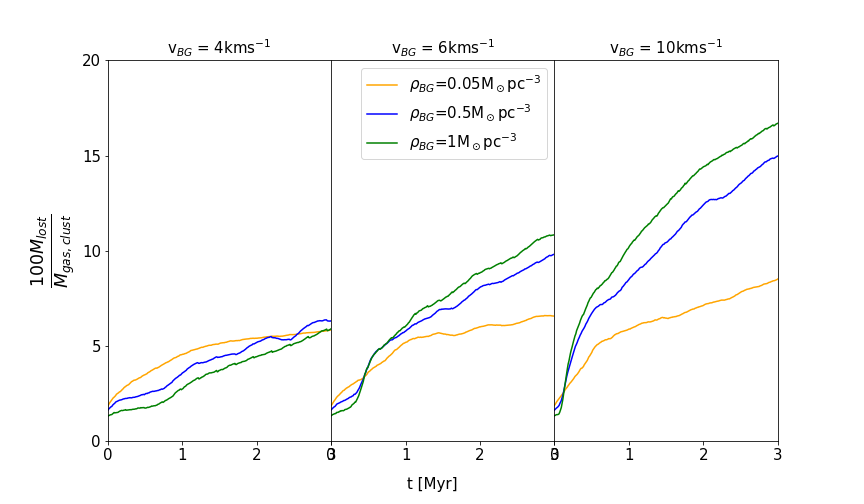}
    \caption{Percent of total initial cluster gas mass that becomes unbound throughout the simulation. Gas that is accreted onto the cluster from the background is not included. The left, middle, and right panels show the simulations of a cluster travelling through an ambient background in which the cluster velocity is 4, 6, and 10kms$^{-1}$ respectively. The colours correspond to the density of the ambient background.}
    \label{fig:PD_tot}
\end{figure*}

\begin{figure}
    \centering
    \includegraphics[scale=0.25]{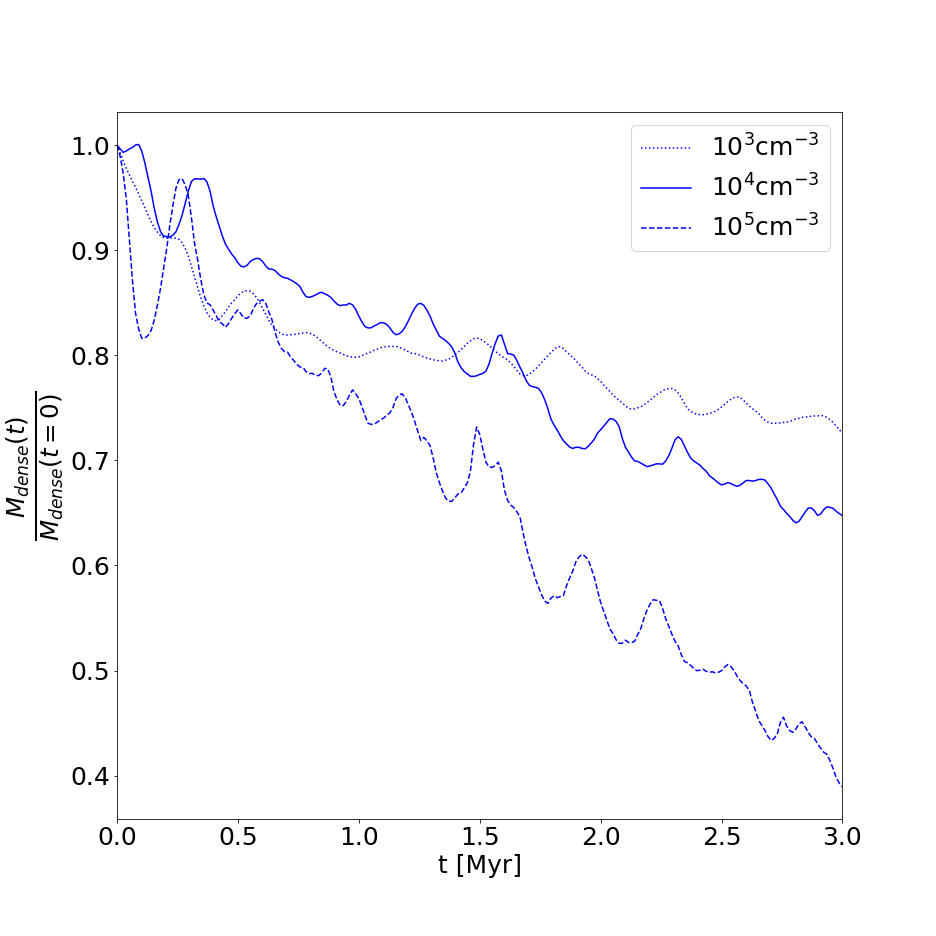}
    \caption{Mass of gas above a given threshold density normalized by that value at the beginning of the simulation for the simulation of a cluster moving through an ambient medium with density $\rho_{\mathrm{BG}}=0.5$M$_\odot$pc$^{-3}$ and a velocity of 4kms$^{-1}$. The dotted, solid and dashed lines correspond to density thresholds of 10$^3$, 10$^4$, and 10$^5$cm$^{-3}$ respectively.}
    \label{fig:SF_dens_streams}
\end{figure}

\subsubsection{Bound Cluster Material}
\label{sec:CtS_BCM}
We begin our analysis of these simulations by looking at the bound cluster material. This helps us better understand how the mass of the cluster is changing as a result of interactions with the background medium. The total cluster mass can change through two ways: accretion of the diffuse background gas onto the cluster, or loss of some of the initially bound cluster mass (stars or gas). To determine the boundedness of any particle in our simulation, we calculate the potential energy (U) felt on a given particle (star or gas) from every other particle in the simulation (stars and gas). We then compare that to the kinetic energy of that particle (T) calculated with respect to the centre of mass of the cluster. Finally, if $T + U < 0$ for a given particle, we consider that particle bound to the cluster. If not, it is unbound.

We find that the boundedness of the stellar component of the cluster remains roughly unchanged for the entire 3Myr in every simulation in table \ref{tab:CtS_sims}. Each simulation loses $\approx$ 1.5 to 2M$_\odot$ of stellar material through unbinding by the end of the simulation which is negligible compared to the total mass of the cluster.

To analyze the gas component, we first calculate the total mass gained by our cluster from the ambient background. We then compare that to the increase in mass of the cluster calculated using the Bondi-Hoyle accretion formalism (\citealt{bondi}, \citealt{shima}) which presents a theoretical accretion rate for a point mass travelling through an ambient medium. An example of these values for the simulation of a cluster moving through an ambient medium with density $\rho_{\mathrm{BG}} = 0.5$M$_\odot$pc$^{-3}$ and velocity 6kms$^{-1}$ can be seen as the blue and black lines respectively in figure \ref{fig:stream_mass2}. We see that using the Bondi-Hoyle formalism underestimates the total mass gained by the cluster likely because our cluster is not a point mass. Varying potential distributions have been shown to have an effect on the Bondi-Hoyle accretion formalism (\citealt{naiman}). We find the same discrepancy for all of our simulations of a cluster moving through an ambient medium implying that Bondi-Hoyle accretion may not be sufficient in describing the amount of mass gained by a cluster as it travels through an ambient medium.

We now compare the total mass accreted onto the cluster to the total bound gas mass of the cluster which considers both accretion from the background medium, and loss of cluster gas through unbinding. We show this total bound gas mass as the orange line in figure \ref{fig:stream_mass2} for the simulation with $\rho_{\mathrm{BG}} = 0.5$M$_\odot$pc$^{-3}$ and $v_{\mathrm{inj}} = 6$kms$^{-1}$. From this plot, we see that the total bound cluster gas mass is decreasing with time. This occurs for all of our simulations with a cluster velocity above 4kms$^{-1}$. However, the amount of mass removed from the cluster throughout this interaction is small compared to the total mass of the cluster. We calculate $M_{\mathrm{lost}}/M_{\mathrm{gas,clust}}$ where $M_{\mathrm{lost}}$ is the total mass of gas that is lost from the cluster, and $M_{\mathrm{gas,clust}}$ is the total cluster gas mass, for this simulation. Here, $M_{\mathrm{gas,clust}}$ includes all the gas that originally belonged to the cluster (column 3 in table \ref{tab:clusts}) and gas mass from the background medium that has become bound to the cluster. We find that only $\approx 5$ percent of the cluster gas mass is stripped off of the cluster from its interaction with the background medium. We perform this comparison for all of our simulations of a cluster moving through an ambient medium and show our results in figure \ref{fig:PD_tot}.

When the background velocity is low, all simulations result in the cluster losing $\approx 5$ per cent of its total gas mass. As we increase the velocity, the background gas density begins to play more of an important role. In the middle panel, we see that the simulation with the highest background density leads to the highest fraction of gas mass lost for that given velocity. This persists when the background velocity is 10kms$^{-1}$. As well, increasing the velocity increases the rate at which the cluster is losing mass which leads to the cluster losing more mass than it gains in the simulations with $v_{\mathrm{BG}} > 4$kms$^{-1}$. However, in all these panels, the amount of gas mass lost remains small compared to the total mass of the cluster. Therefore, we conclude that movement through a background medium can result in a net decrease in the total mass of a cluster, but that decrease is small compared to the total mass of the cluster.

This analysis also helps us learn more about the accuracy of the sink particle prescription as it would apply to our simulations. Because sink particles do not allow for material to be removed from a sink after the sink has accreted it, our simulations show that they can be overestimating the total mass of the cluster they represent. However, because the amount of mass lost from the cluster is small compared to the total mass of the cluster, this overestimation would only be by a small fraction ($< 20$ per cent).

\begin{figure*}
    \centering
    \includegraphics[scale=0.3]{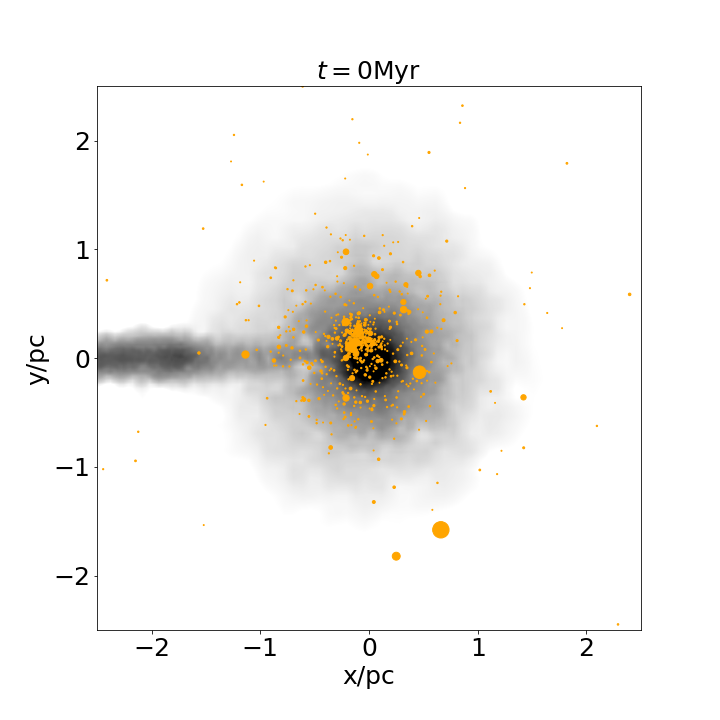}
    \includegraphics[scale=0.3]{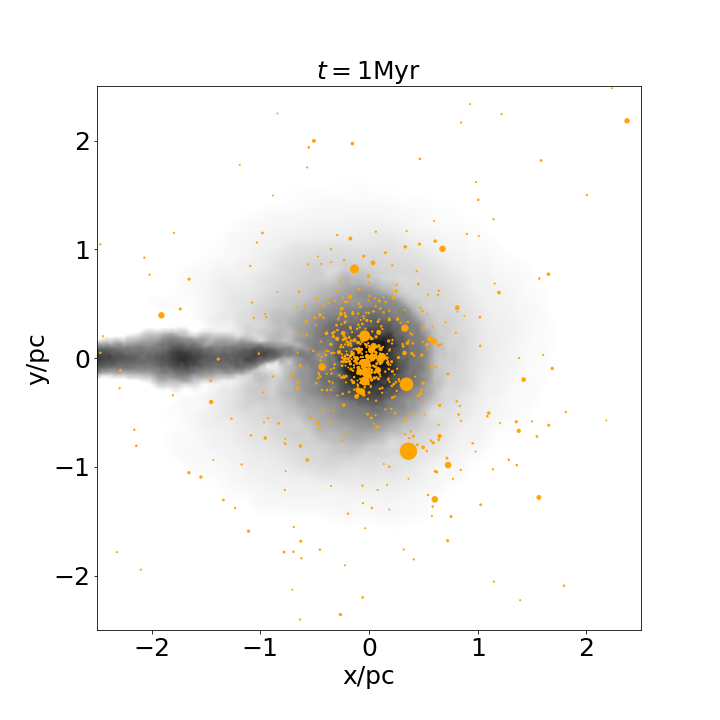}
    \includegraphics[scale=0.3]{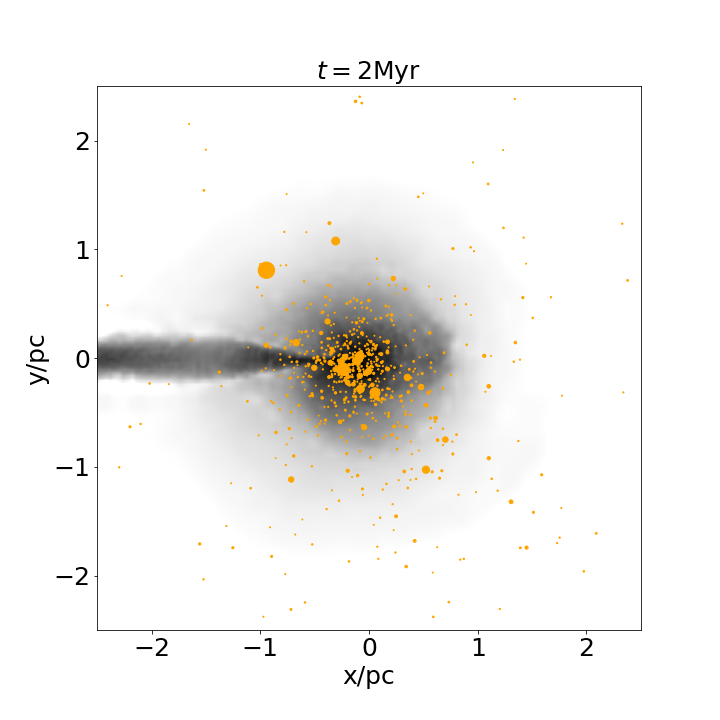}
    \includegraphics[scale=0.3]{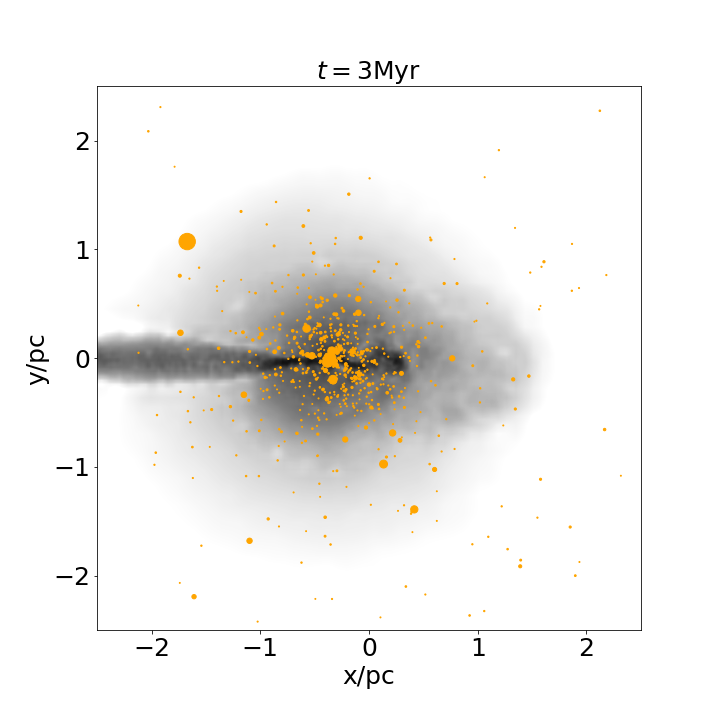}
    \caption{Snapshots of the stars and gas from filamentary accretion run FwC1250. The orange circles represent the stars from the cluster with their size scaled to the mass of the star. Gas is shown in greyscale with darker regions showing higher density gas. The minimum density shown is 10M$_\odot$pc$^{3}$ and the maximum density shown is 10$^4$M$_\odot$pc$^{-3}$.}
    \label{fig:fwc1200_snap}
\end{figure*}

\begin{figure}
    \centering
    \includegraphics[scale=0.25]{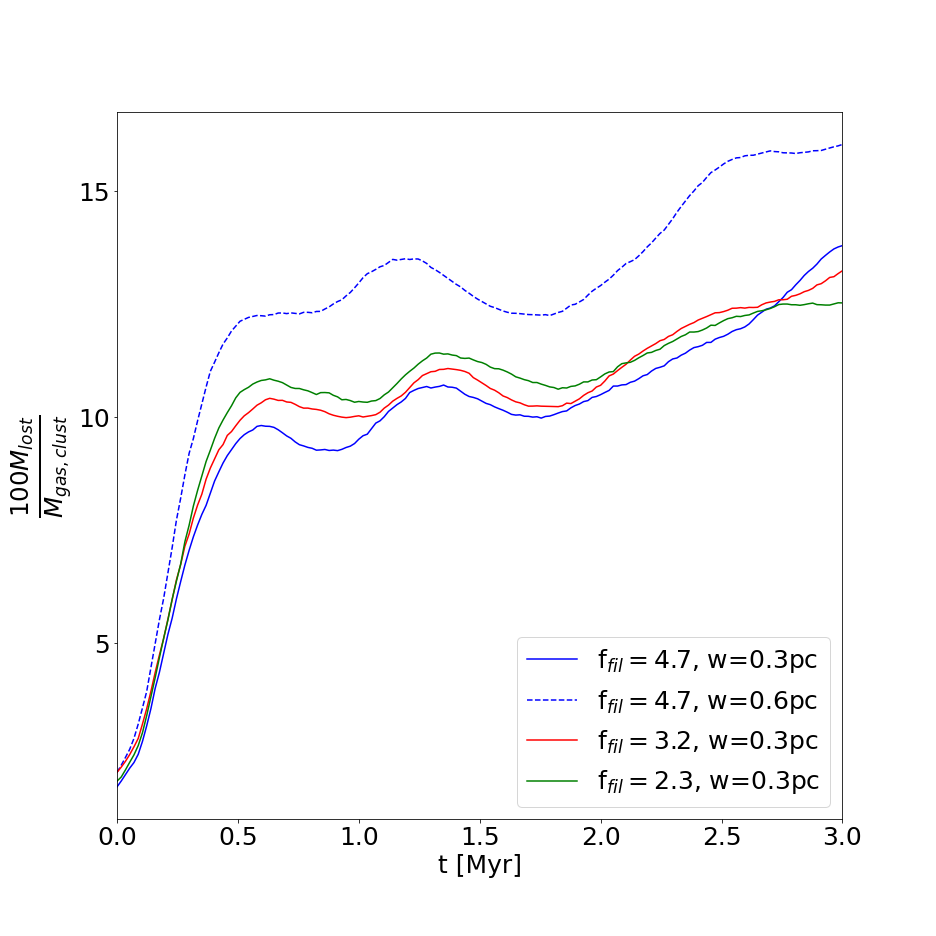}
    \caption{Percent of total cluster gas mass that is unbound throughout the simulation for all simulations of a single filament impacting our smaller cluster. Colour describes $f_{\mathrm{fil}} \propto \rho w^2$ of the filament in a given simulation and the dashed line denotes a larger filament width.}
    \label{fig:single_fils_bound_gas}
\end{figure}

\subsubsection{Star Forming Gas}
\label{sec:CtS_SF}

Sink particles in the H18 simulations convert gas into stars with a constant efficiency of 20 per cent every free fall time. While our simulations do not include a star formation prescription, we can trace the amount of dense gas present to understand what effect movement through a background medium would have on the star formation efficiency of our clusters. We consider gas with densities above $10^4$ and $10^5$cm$^{-3}$ which are quoted density ranges above which stars begin to form (e.g \citealt{evans09}, \citealt{heid10}, \citealt{lada2010}, \citealt{lada2012}). As well, the former is the threshold density above which sinks are formed in the H18 simulations. We also consider gas above $10^3$cm$^{-3}$ as this has been used as sink particle formation thresholds in previous, larger scale simulations as well (e.g. \citealt{dobbs_1}).

We find that our cluster loses $\approx 30$ per cent of its mass of gas above 10$^4$cm$^{-3}$ by the end of the simulation regardless of the density of the ambient background gas. An example of this can be seen as the solid line in figure \ref{fig:SF_dens_streams} for the simulation with a background density of $\rho_{\mathrm{BG}} = 0.5$M$_\odot$pc$^{-3}$ and gas velocity of 4kms$^{-1}$. This consistency rises from the fact that, as the cluster moves through the diffuse medium, the outer layers are those that are most perturbed while the cluster core remains mostly unperturbed due to its higher potential. We find the same slow decrease when looking at the total mass of gas with densities above the other two thresholds regardless of the background density. The difference is that we see a decrease of $\approx 25$ and $\approx 50$ per cent by the ends of the simulations for gas with densities above 10$^{3}$ and 10$^{5}$cm$^{-3}$ respectively. As well, we find that these trends do not change when we change the density of the ambient background gas.

We find that the velocity of the ambient medium has no affect on the dense gas present at any time for any of the threshold densities we considered. Increasing the velocity leads to an increase in gas which becomes unbound from the cluster, and this gas is mostly located in the outer regions of the cluster meaning the dense core remains mostly unaffected.

Therefore, we find that, for the values we considered, the density of the ambient background, and the velocity of the cluster travelling through a uniform density ambient background do not play a role in determining the amount of dense gas present in the cluster. In our simulations, we would see less star formation over time as the cluster travels through the ambient medium. This may change when a more realistic background density distribution is considered. We discuss this more in section \ref{sec:discussion}.

\subsection{Single Filaments}

\label{sec:single_fils}

We now move on to a discussion of the accretion from a single filament onto our smaller star cluster. These correspond to the simulations FwC1250, FwC850, FwC600, and FwC310 in table \ref{tab:fil_sims}. These simulations provide us with a baseline to which we can compare our simulations of multiple filament systems. Each of these four simulations have different filament densities $\rho_0 = 1250, 850, 600, $ and $310$M$_\odot$pc$^{-3}$ with filament widths $w = 0.3, 0.3, 0.3, $ and $0.6$pc respectively. These correspond to $f_{\mathrm{fil}} = 4.7, 3.2, 2.3,$ and $4.7$. The density and width of the last of these simulations was chosen such that its $f_{\mathrm{fil}}$ is the same as the first. This helps us understand whether the line mass (or accretion rate $M_{\mathrm{line}}v_{\mathrm{fil}}$) of a filament is sufficient in predicting the effect it has on an accreting cluster. The cluster is the same as that from section \ref{sec:CtS}. We show snapshots of the simulation with the higher density filament ($\rho=1250$M$_\odot$pc$^{-3}$) in figure \ref{fig:fwc1200_snap}. The gas density is illustrated by the grey scale with a minimum of 10M$_\odot$pc$^{-3}$ and a maximum of $10^4$M$_\odot$pc$^{-3}$. The orange circles represent the stars in the cluster and their size is scaled with the mass of the star. 

At $t=0$Myr, the filament of gas is seen on the left side of the cluster with a velocity towards the cluster centre. As the filament impacts the core of the cluster, the gas density in the inner region of the cluster starts to decrease. By the end of the simulation, the cluster core has become so diffuse that gas coming from the filament pushes through the cluster and comes out the other side (see dense region on right hand side of cluster in bottom right panel of figure \ref{fig:fwc1200_snap}). We see similar behaviour for all four of these single filament simulations. Energy from the impact of the filament onto the cluster causes the gas component to expand.

The stellar component responds to this accretion by expanding slightly because of the expansion of the gas component. The stellar component core radius grows by a factor of 2 for the three simulations whose filaments have a width of 0.3pc and grows by a factor 3 for the simulation whose filament width is 0.6pc.

\subsubsection{Bound Material}
\label{sec:bound_fils}

First, we analyze how gas accretion from a single filament affects the total mass of the cluster by looking at the bound mass of the stellar and gas components of our cluster as it accretes gas from the filament.

We find that the boundedness of the stellar component of the cluster for these four single filament simulations is not very affected by the filamentary accretion. Throughout the entire simulation, less than 2M$_\odot$ of stellar mass become unbound from the cluster for each simulation similar to the simulations discussed in section \ref{sec:CtS_BCM}. 

Conversely, the total mass of the gas component of the cluster is strongly affected. The cluster is constantly gaining gas mass from interactions with the flowing filament. We find that $\approx 90$ percent of the total mass of the filament is bound to the cluster for all four of these simulations. We also find that accretion from a filament causes a fraction of the gas originally belonging to the cluster to become unbound. To understand the effect this has on the total cluster gas mass, we calculate the fraction of the total cluster gas mass that is unbound from the cluster as a function of time. We show this as a percentage for these four single filament simulations in figure \ref{fig:single_fils_bound_gas}. From this plot, we see that the amount of gas mass which becomes unbound from the cluster is small compared to the total mass of the cluster. Therefore, the net effect of accretion from a filament is an increase in the mass of the cluster. Furthermore, this trend is similar for all simulations with a smaller filament width regardless of the density we considered. While a larger filament width results in more gas mass becoming unbound from the cluster, the amount remains small compared to the total mass of the cluster. 

We conclude that our cluster loses a small fraction of its mass from filamentary accretion, and that this fraction is similar for all filament widths and densities considered. This result has implications for the sink particle prescription as well. Because sinks do not allow for the loss of material, and we find that this loss is small compared to the total mass of the cluster, we can conclude that sink particles are only slightly overestimating the total mass of the cluster they represent when considering accretion from a single filament.

\subsubsection{Star Forming Gas}
\label{sec:SFG_fil}
We now track the total mass of dense gas in these simulations to help us understand how accretion from a single filament affects star formation inside the central cluster.

We find that the total mass of gas with densities above 10$^4$cm$^{-3}$ slowly decreases as the the cluster is impacted by the filament for each of these four simulations. By the end of the simulation, this decrease is only by $\approx 40$ per cent for the simulations whose filaments have a width of 0.3pc. Conversely, the simulation with a filament of width 0.6pc sees a much more drastic decrease in dense gas mass, namely, by around 80 per cent by the end of the simulation. This trend persists when looking at densities above 10$^5$cm$^{-3}$ as well. However, in this case, the total mass of dense gas decreases by $> 90$ per cent for all four of these simulations by the end of the simulation. Lastly, we find that the mass of gas above 10$^3$cm$^{-3}$ plateaus for the simulations which contains filaments of width 0.3pc and have densities of 1250 and 850M$_\odot$pc$^{-3}$. The remaining two simulations see consistent decrease in gas above 10$^3$cm$^{-3}$ with time.

Therefore, for our single filament simulations, we find that the total mass of potentially star forming dense gas present as the simulation evolves is not dependent on the line mass of the filament that is impacting the cluster. It is better to consider the density and width of the filament separately. We also conclude that the amount of potentially star forming gas in these simulations can either remain constant or decrease depending on the threshold chosen above which star formation is allowed. We do not see any net increase in the amount of potentially star forming gas in these single filament simulations.

\subsection{Double Filaments}
\label{sec:2fil}
\begin{figure*}
    \centering
    \includegraphics[scale=0.3]{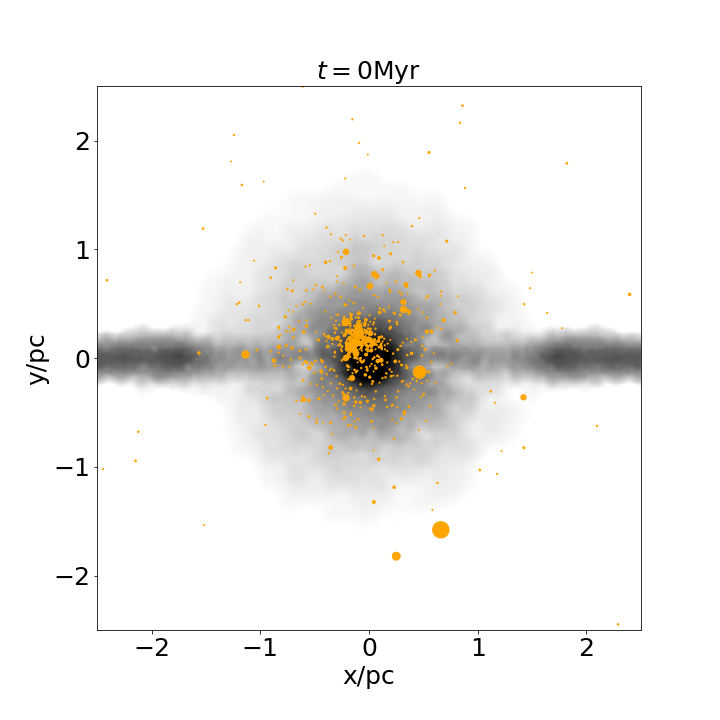}
    \includegraphics[scale=0.3]{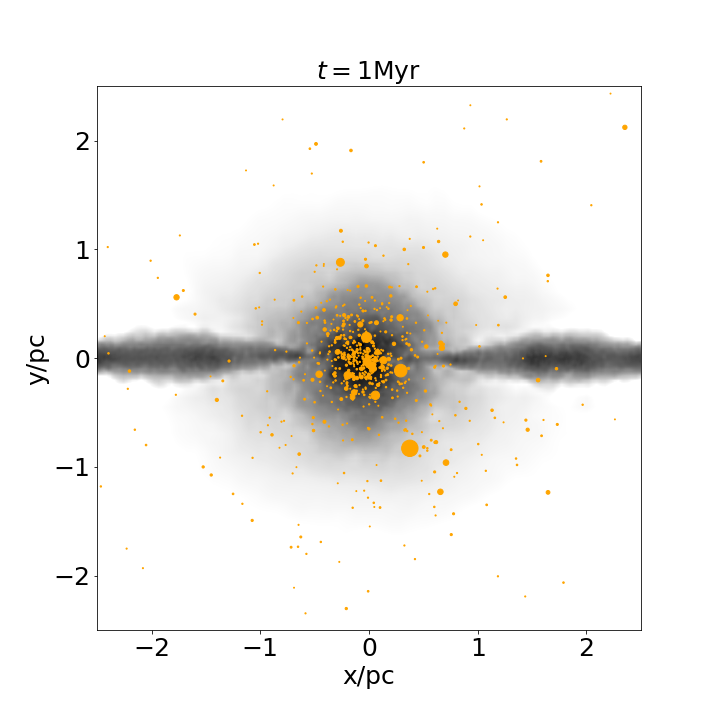}
    \includegraphics[scale=0.3]{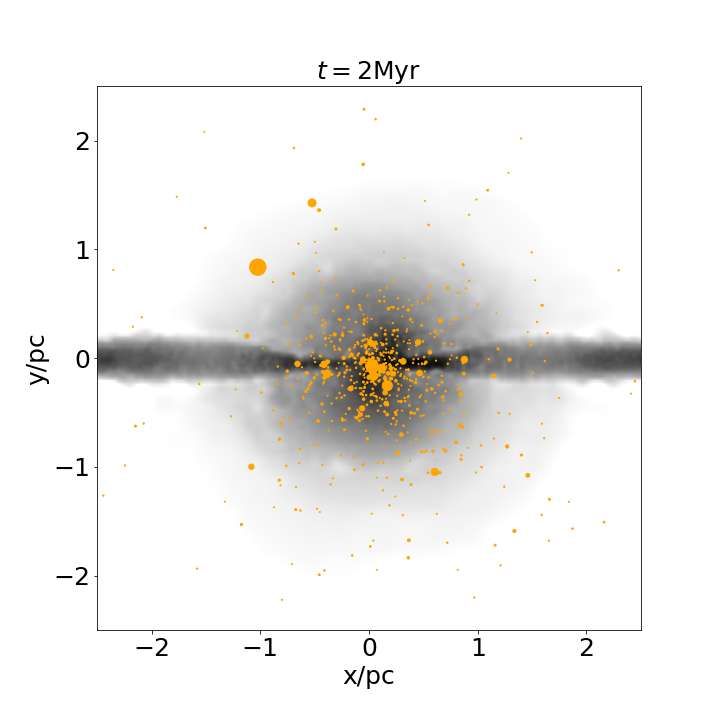}
    \includegraphics[scale=0.3]{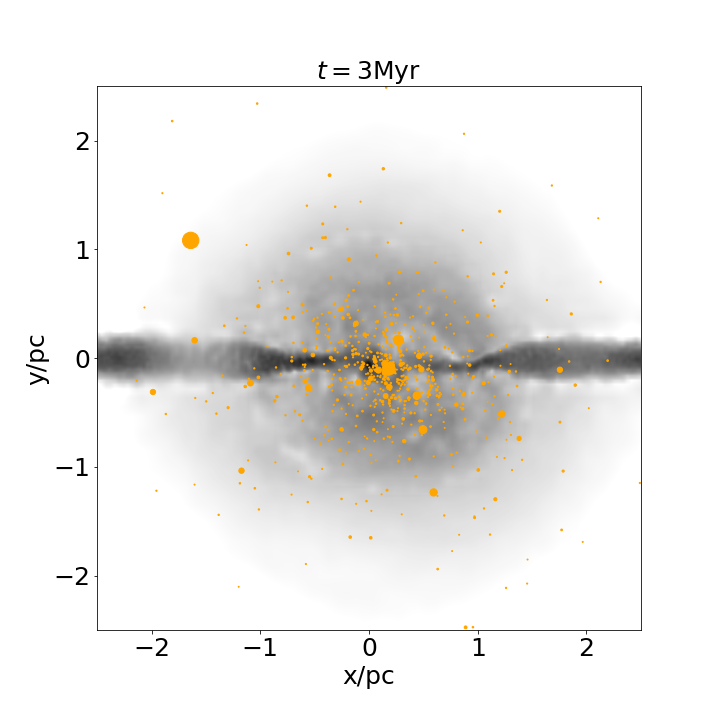}
    \caption{Same as figure \ref{fig:fwc1200_snap} but for the simulation 2FwC1250.}
    \label{fig:2fwc1200_snap}
\end{figure*}

\label{sec:SFG_2fil}
\begin{figure}
    \centering
    \includegraphics[scale=0.25]{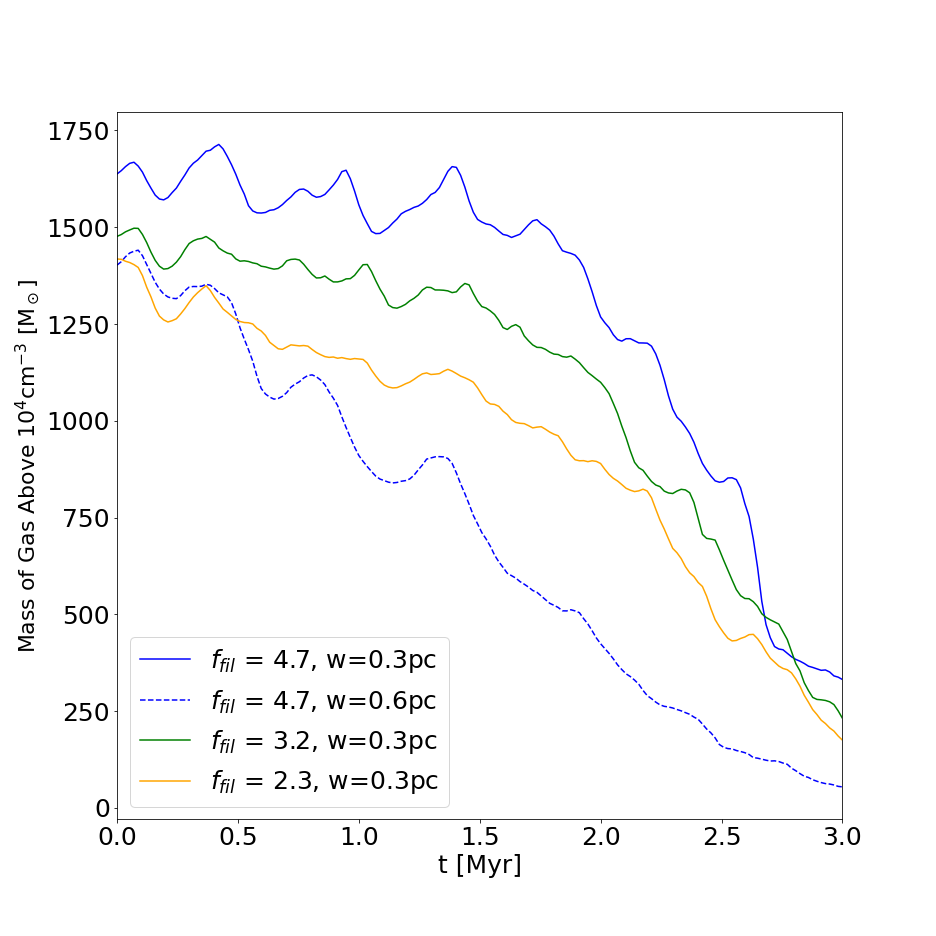}
    \caption{Total mass of gas above 10$^4$cm$^{-3}$ for simulations of two filaments accreting onto our smaller star cluster. Colour describes $f_{\mathrm{fil}} \propto \rho w^2$ of the filament in a given simulation and the dashed line denotes a larger filament width.}
    \label{fig:SF_thresh2}
\end{figure}

We now turn our discussion towards the simulations in our suite of accretion of gas from two parallel filaments onto our smaller star cluster. These simulations are more representative of the real universe than those performed using a single filament. This is a consequence of the formation mechanism of prestellar cores which sees them forming within dense filaments, and not on the edges (\citealt{andre_SF}). These correspond to simulations 2FwC1250, 2FwC850, 2FwC600, and 2FwC310 in table \ref{tab:fil_sims}. The densities and width of these filaments are the same as those from our single filament simulations discussed in section \ref{sec:single_fils}, but we have doubled the single filament and mirrored it on the other side of the cluster. As well, the central star cluster is the same as in our single filament simulations. Snapshots from the double filament simulation with filaments of density 1250M$_\odot$pc$^{3}$ and width 0.3pc can be seen in figure \ref{fig:2fwc1200_snap}. 

We see that impact onto the cluster from filamentary gas leads to expansion of the clusters gas component. In these simulations, the mass of new dense gas accreted onto the cluster core from the two filaments is sufficient enough to prevent the expansion of the stellar component found in the single filament simulations. 

\subsubsection{Bound Material}
\label{sec:2fil_bound}
We first look at the amount of material that is bound to the cluster as it accretes filamentary gas in these double filament simulations. Similarly to the single filament simulations, a negligible amount of stellar mass is unbound from the cluster in all of these simulations ($\approx 2$M$_\odot$ at most). 

However, the total mass of the gas component of the cluster increases more in the double filament simulations than in the single filament simulations, as expected. The increase in the gas mass of a cluster accreting from two filaments of a given density is roughly twice that of cluster accreting from one filament of the same width and density. The mass of gas removed from the cluster, however, does not scale this way. We find that roughly the same amount of mass is removed from the cluster in these simulations as those of accretion from a single filament. As a consequence of this, the fraction of the total cluster gas mass that becomes unbound is slightly less for these simulations than those from figure \ref{fig:single_fils_bound_gas}, but not substantially. We find that this percentage lies in the same range as those in figure \ref{fig:single_fils_bound_gas}. As well, in these double filament simulations, there is no substantial difference between the fraction of mass that becomes unbound for different filament parameters, similar to section \ref{sec:bound_fils}.

This result also shows that the magnitude of the overestimation made by the sink particle prescription in calculating the total gas mass of the cluster is roughly equal between accretion from one or two filaments in our suite. We discuss the implications of this further in section \ref{sec:discussion}.

\subsubsection{Star Forming Gas}

Similar to section \ref{sec:SFG_fil}, we analyze the total mass of dense gas above 10$^3$, 10$^4$ and 10$^5$cm$^{-3}$ in our double filament simulations. We show the total mass in gas with densities above the 10$^4$cm$^{-3}$ threshold for these four double filament simulations in figure \ref{fig:SF_thresh2}. Our simulations involving the two densest filaments (blue and green line in figure \ref{fig:SF_thresh2}) contain a roughly constant amount of mass above 10$^4$cm$^{-3}$ for $\approx 1.5$Myr. This is in contrast to the simulations containing a single filament where we saw steady decrease in gas above 10$^4$cm$^{-3}$ regardless of filament parameters. The second filament can act as another source of dense gas to help prevent the gas core of the cluster from becoming too diffuse throughout the simulation. However, similar to the dense gas present in the single filament simulations, wider and more diffuse filaments result in a fast decrease in the total mass of gas above 10$^4$cm$^{-3}$.

Regarding gas above 10$^5$cm$^{-3}$, we see the same trend in these simulations as we did in our single filament simulations. However, when we consider the 10$^{3}$cm$^{-3}$ threshold in these double filament simulations, we find that the total mass of gas above this density increases over time when the filament density is high (1250 or 850M$_\odot$pc$^{3}$) by $\approx 5$ per cent in 3Myr. Conversely, when the filament density is low (600 or 310M$_\odot$pc$^{3}$), the total mass of gas above this threshold plateaus and remains roughly constant for most of the simulation.

Therefore, it is important to know both the width and density of a filament to better understand its impact on the total mass of star forming gas in the accreting cluster when two filaments are accreting. As well, accretion from two filaments has the potential to increase the amount of dense gas available for star formation over time in our simulations, unlike accretion from a single filament.

\subsection{Dependance of Response to Filamentary Accretion on Cluster Parameters}
\label{sec:other_clust}
We performed all of our single and double filament simulations with a more massive central cluster (see simulations that end in C2 in table \ref{tab:fil_sims}). This allows us to determine if the trends found in the previous sections are ubiquitous across star clusters. The cluster in these simulations has a similar gas mass ($M_g \approx 3\times10^4$M$_\odot$) as the previous cluster but an increased stellar mass ($M_s \approx 3\times10^4$M$_\odot$) leading to an increased total mass. The filament setups and orientations are the same as the corresponding simulations with the less massive central cluster.

We begin with a discussion of the amount of bound material present in simulations of single and double filamentary systems accreting gas onto our more massive cluster. Beginning with the stellar component, all eight simulations involving the more massive cluster lose less than 1 per cent of stellar mass by their completion. This is independent of any filament parameter including density, width, and number of filaments and is a negligible fraction of the total mass of the cluster. Regarding the gas component, we perform the same analysis as in the previous sections and reach similar conclusions. We find that the central cluster loses a small percentage of its total mass due to interactions with filaments and that this percentage lies in the same range as those from sections \ref{sec:bound_fils} and \ref{sec:2fil_bound}. We also find that this is independent of number of filaments, filament width, or filament density.

We now discuss the potentially star forming dense gas present in these simulations. We find that the width of the filament(s) interacting with the central cluster plays less of a role in affecting the amount of dense gas present in the simulations. When considering gas above 10$^4$cm$^{-3}$ in the simulations with our widest filament ($w = 0.6$pc), we do not see the stark decrease in potentially star forming gas that we did in figure \ref{fig:SF_thresh2} for example. Rather, the total mass of gas above 10$^4$cm$^{-3}$ in these simulations follow a similar trend to that in simulations with lower filament width. Therefore, the size of a filament with respect to the total size of the central cluster is important in determining the amount of potentially star forming gas present throughout the simulation for gas above 10$^4$cm$^{-3}$. For gas above 10$^3$ and 10$^5$cm$^{-3}$, we find very similar variance between different filament setups as those found in our simulations with a smaller central cluster showing that this dependence is present for all density thresholds we considered. 

Therefore, we conclude that the change in total mass and size of the clusters we considered is important when considering the fraction of dense gas available for star formation in our clusters. It does not have an affect on the fraction of the total mass of the cluster that becomes unbound.

\begin{figure*}
    \centering
    \includegraphics[scale=0.3]{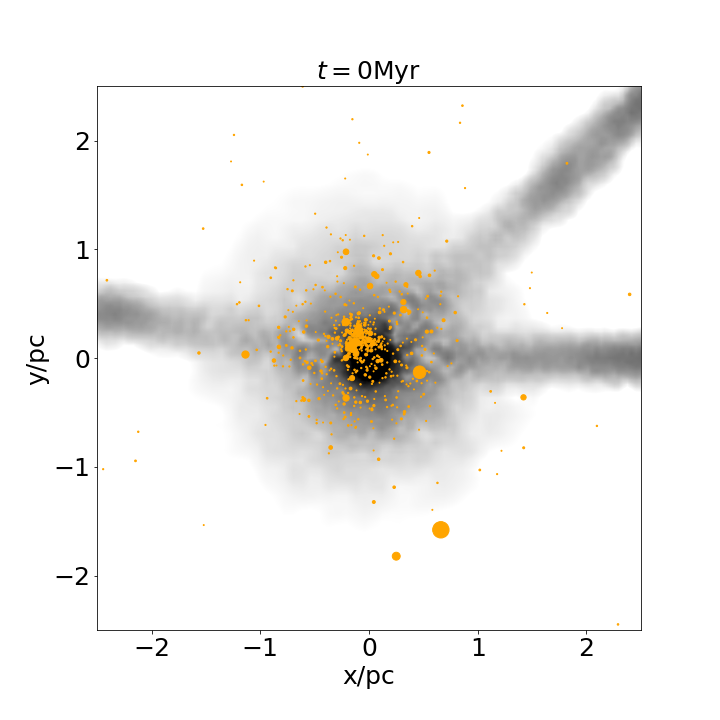}
    \includegraphics[scale=0.3]{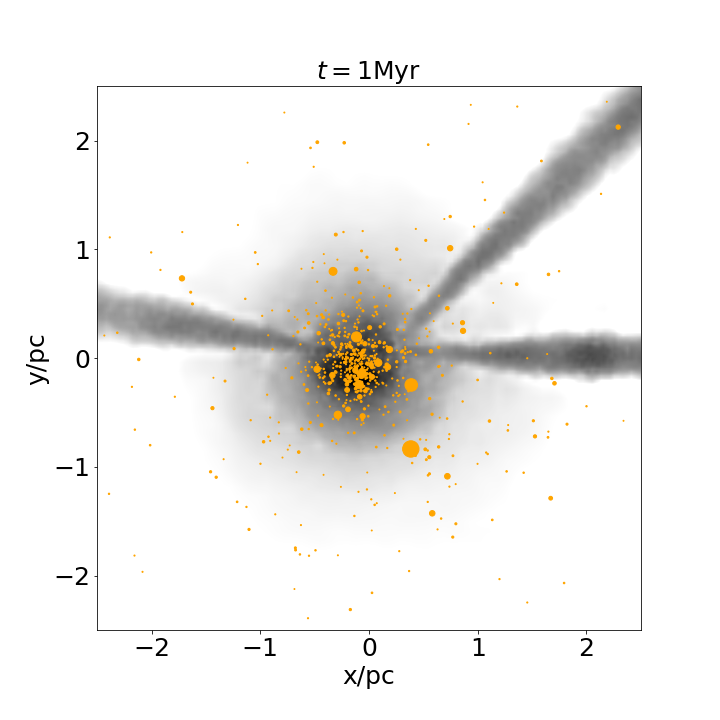}
    \includegraphics[scale=0.3]{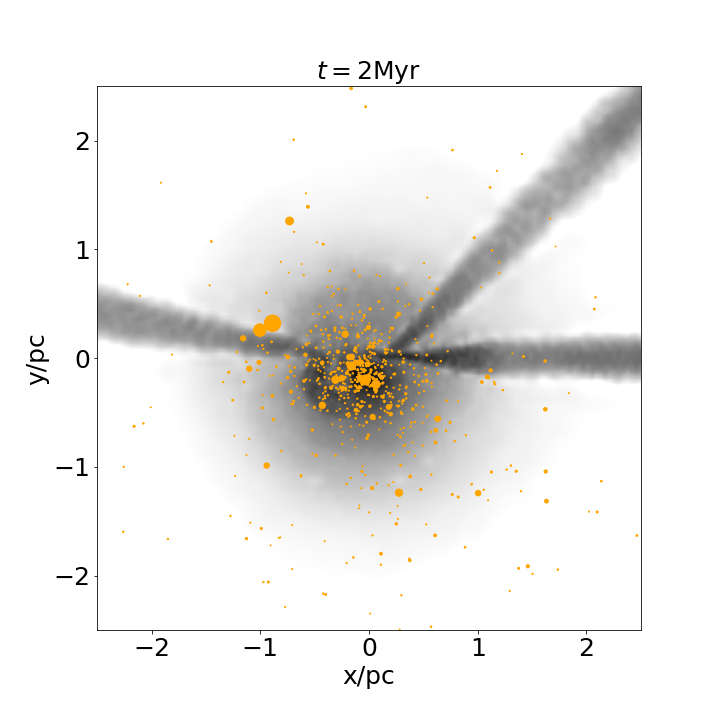}
    \includegraphics[scale=0.3]{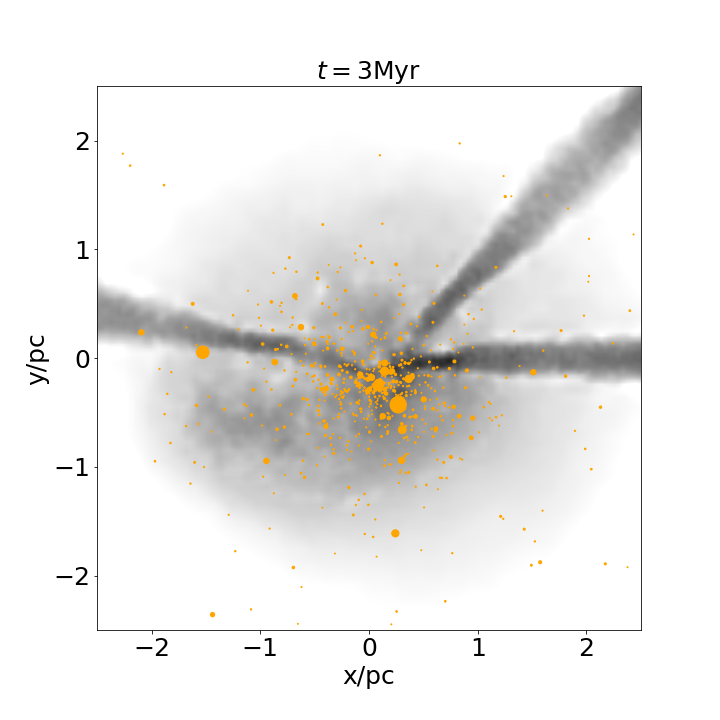}
    \caption{Same as figure \ref{fig:fwc1200_snap} but for the simulation SDC.}
    \label{fig:SDC_snap}
\end{figure*}

\begin{figure}
    \centering
    \includegraphics[scale=0.25]{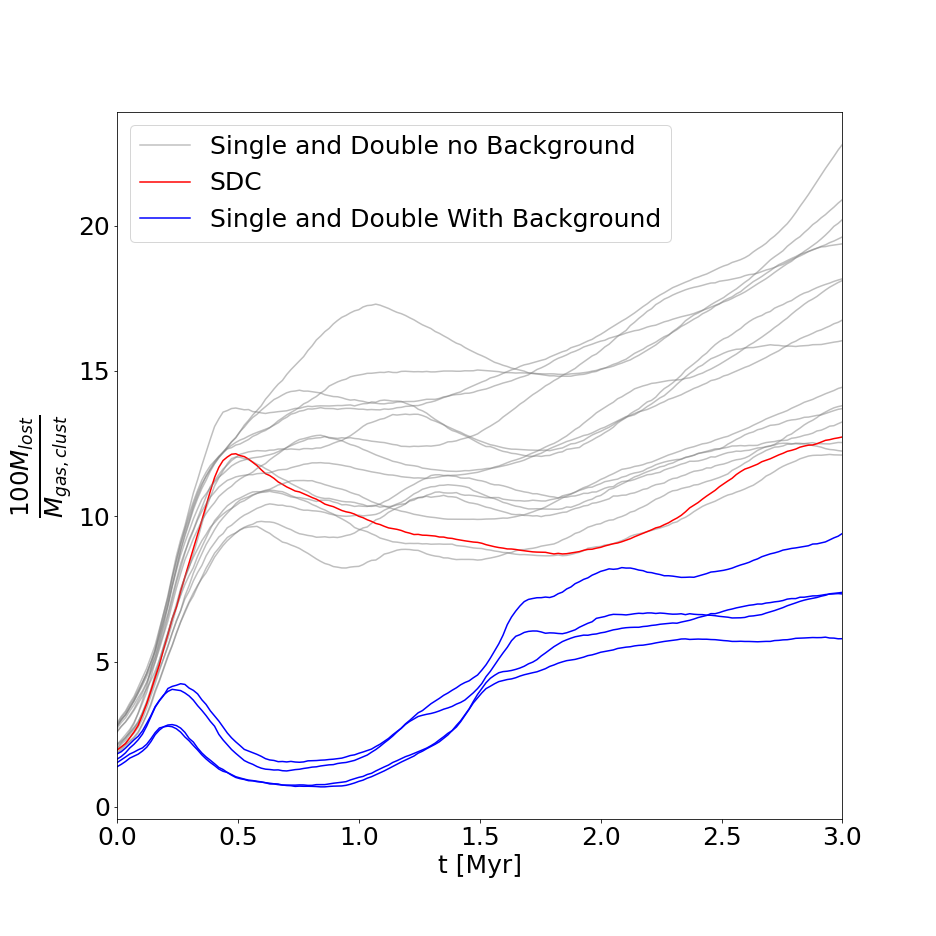}
    \caption{Same as figure \ref{fig:single_fils_bound_gas} but for all filament simulations. The grey lines correspond to all single and double filament simulations with no ambient background gas (sections \ref{sec:single_fils}, \ref{sec:2fil}, and \ref{sec:other_clust}), the red line shows the SDC simulation (section \ref{sec:SDC}), and the blue lines correspond to single and double filament simulations that include a background medium (section \ref{sec:FCS}).}
    \label{fig:all_PD_fils}
\end{figure}

\subsection{SDC13}
\label{sec:SDC}
A common pattern present in many star forming regions is a distribution of filaments in a "Y" shape involving three main filaments feeding their gas into a central, often times star forming, region (e.g. \citealt{koy}, \citealt{andre_2010}). This pattern is also present in simulations of GMCs (e.g. \citealt{ekster}, \citealt{starforge2}). One commonly observed region that contains this filament distribution is the SDC13 region. SDC13 is a hub filament infrared dark cloud located in the galactic plane at a distance of 3.6$\pm$0.4kpc (\citealt{SDC}) and its filamentary network is distributed in a "Y" shape with the filaments feeding gas into a central star forming region. The formation of this particular filament distribution has been studied by \citet{SDC_2} and its subsequent evolution along with the orientation of the filaments has been studied in \citet{jvla}. We take these filament orientations and create our own simulation of a central cluster accreting dense gas in a similar manner to the SDC13 region. We show snapshots of this simulation in figure \ref{fig:SDC_snap}. The parameters of the filaments in this simulation are shown as run SDC in table \ref{tab:fil_sims}. The densities and widths of the simulations used match observations of the SDC13 filaments from \citet{SDC}.

This simulation behaves similarly to our double filament accretion simulation that contains higher density filaments (2FwC1250). The cluster loses very little stellar mass, and it loses a similar amount of gas mass as all previous filament accretion simulations. This can be seen as the red line in figure \ref{fig:all_PD_fils} where we show the percentage of the total cluster gas mass that is lost from accretion from filaments for all simulations in table \ref{tab:fil_sims}. As well, we find the same trends as in 2FwC1250 regarding gas above the three density thresholds analyzed in section \ref{sec:SFG_2fil}. We discuss this simulation in comparison to all other filament simulations in section \ref{sec:discussion}.

This "Y" shape filament configuration is one of many configurations observed throughout star forming regions. For example, hub filament systems can contain very complex filament configurations with many more than three filaments accreting onto a star forming region. We discuss such complexities and how they connect to our work further in section \ref{sec:discussion}.

\subsection{Filamentary Accretion With Ambient Background Gas}
\label{sec:FCS}
Lastly, in this section, we discuss the simulations FCS1250\_05, FCS1250\_1, 2FCS1250\_1, and 2FCS1250\_05 (see table \ref{tab:fil_sims}) which are mixtures of our ambient medium and filament simulations. These simulations better resemble the environments found in the real universe where a cluster has formed along a filament that is surrounded by less dense GMC gas and is accreting the surrounding gas. The first two comprise of a single filament with density $\rho = 1250$M$_\odot$pc$^{-3}$ and an ambient background medium with density $\rho_{\mathrm{BG}} = 0.5$ (FCS1250\_05) or 1M$_\odot$pc$^{-3}$ (FCS1250\_1) filling the simulation box. The latter two are the same but with two filaments of density $\rho = 1250$M$_\odot$pc$^{-3}$ that are set up in the same way as those in section \ref{sec:2fil}. In all four of these simulations, only the filamentary gas is given a velocity towards the cluster. As well, in all four of these simulation, we use the less massive cluster from table \ref{tab:clusts} as our accreting cluster. 

We first look at the amount of bound material present in these four simulations. Similarly to all simulations thus far, the stellar component of each of these simulations does not lose any significant amount of mass through unbinding. Only $< 1$M$_\odot$ of stellar material is unbound by the end of the simulations. As well, we find that the cluster in these simulations loses much less gas mass than that in the simulations of filamentary accretion without a diffuse background. We show the percentage of the total cluster gas mass that is removed from the cluster for these four simulations as the blue lines in figure \ref{fig:all_PD_fils}. We also find that there is little difference between one filament and two filament accretion for a given background density regarding the amount of unbound gas present. Therefore, a background medium can be used to help lower the amount of gas removed from a cluster as it interacts with a filament. This implies that a background medium can help the sink particle prescription better represent star clusters accreting gas from filaments.

We now move on to a discussion of the amount of potentially star forming, dense gas present in these simulations. Beginning with the simulations involving accretion from a single filament, we find that gas above 10$^3$cm$^{-3}$ stays constant for the entirety of the simulation regardless of the background ambient density. The total mass of gas above 10$^4$cm$^{-3}$ stays constant for $\approx$ 1Myr which is contrary to the simulations with one filament and no ambient background (see section \ref{sec:SFG_fil}). This implies that the added pressure provided by an ambient medium can help stabilize the amount star forming gas in our single filament simulations, but only for a couple free-fall times. Lastly, gas above 10$^{5}$cm$^{-3}$ decreases consistently throughout both simulations. 

When we look at the two simulations with two filaments and different ambient background densities, we find that gas above 10$^3$cm$^{-3}$ increases by 10 per cent for both simulations. The mass of gas above 10$^4$cm$^{-3}$ and 10$^5$cm$^{-3}$ stays constant in both simulations for $\approx 2$Myr and $\approx 1$Myr respectively. Therefore, we conclude that the addition of a background medium does not result in a substantial change with regards to the amount of dense gas present in our filamentary accretion simulations.

\section{Summary and Discussion}
\label{sec:discussion}

We have run simulations of gas accretion from both the ambient background medium, and dense filaments onto star clusters whose parameters we obtained from sink particles in a large scale GMC star cluster formation simulation (\citealt{Howard2018} (H18)). To simulate accretion from the ambient background medium, we used observations and simulations of GMCs to choose three different background gas densities through which our cluster travelled with three different velocities. To simulate accretion from filaments, we used observations of filamentary regions such as SDC13 to obtain realistic filament densities, widths, and orientations. Throughout both simulation suites, we found that $\lessapprox 20$ per cent of the total gas mass of the cluster is removed from the cluster. Furthermore, this is independent of filament setups (number and orientation) or parameters (density and width) we considered in this work. This has applications to the sink particle prescription used in the H18 simulations. Namely, that the mass quoted by a sink particle representation of our simulations would provide an upper limit to the actual mass of the clusters in our simulations. We also found that the amount of gas above star forming densities inside a cluster can be kept constant from accretion from two filaments if the filaments are dense enough and if their width is small compared to the size of the cluster. 

A boundedness check to determine gas accretion onto sinks like that used in H18 is not the only accretion prescription used by sink particles. Some prescriptions, such as that put forward by \citet{krumholz_sink} and subsequently used in the \texttt{RAMSES} N-body and SPH code (\citealt{new_sink}) use idealized Bondi-Hoyle accretion models to calculate accretion rates onto sink particles. Our work shows that, in cases where these sink particles are meant to represent young, gas-rich star clusters, such accretion formalisms may lead to different inaccuracies regarding the mass of the clusters than the H18 sink particle prescription.

Our suite of simulations of a cluster moving through an ambient medium involve a very idealized treatment for the background gas -- a constant density. This was done to test idealized models (Bondi-Hoyle) against more realistic accretion scenarios. In reality, a single GMC can have a wide range of densities through which a young subcluster can travel (\citealt{chevance_2}). To accurately include this in our simulations would involve zooming into regions directly from GMC simulations such as H18 to get a more realistic distribution for the background gas. Such a realistic distribution would include regions of varying density implying that the travelling cluster would be interacting with under and over densities. Therefore, the resulting change in the mass of the cluster would be more complicated than in our current simulations. We will address this in future work.

To compare results from our filament simulations to observations, we consider measured filamentary accretion rates. Observations show that velocity gradients along filaments are useful in helping young star clusters grow in mass (e.g. \citealt{lee}, \citealt{yuan}, \citealt{chen_fil}). The corresponding accretion rate of a given filament onto such a young cluster can be calculated using the line mass of that filament. It is defined as $\dot{M}_\mathrm{acc} = M_{\mathrm{line}}v_{\mathrm{inj}}$ where $\dot{M}_\mathrm{acc}$ is the accretion rate, $M_{\mathrm{line}}$ is the line mass of the filament, and $v_{\mathrm{inj}}$ is the velocity of the filament (\citealt{kirk}). From velocity gradient and mass observations of filaments, one can therefore conclude the accretion rate given a projection angle through assuming the above simple cylindrical model. This is an idealized cylindrical model that assumes that all of the mass being funneled to the cluster by the filament is bound. We can compare this accretion rate to the rate of change of bound mass of the cluster to see how accurately the accretion rate represents the total change in the clusters mass. 

We find that, for all of our single and double filament simulations involving our smaller cluster, the rate of change of the cluster mass averages at $\approx 0.7\dot{M}_\mathrm{acc}$. For the simulations involving our more massive central cluster, this value averages at $\approx 0.5\dot{M}_\mathrm{acc}$ and is similar for all filament setups and parameters. As well, the introduction of an ambient medium increases the rate of change of the cluster mass to $\approx 0.8\dot{M}_\mathrm{acc}$. Lastly, our SDC13 simulation results in an accretion rate that plateaus at $\approx 1\dot{M}_\mathrm{acc}$ due to the increased potential from the gas funneled by all three filaments. Therefore, idealized cylindrical models represent the total mass of our simulations best when an ambient background is included, or when the common "Y" shape is used for our filament orientations. In our other simulations, they are overestimating the total mass of the cluster.

All of our filament simulations inject filamentary gas towards the cluster at the same velocity (2kms$^{-1}$). This choice was made so that the filament would interact with the central cluster before it had a chance to form dense, potentially star forming cores. This allowed us to explore the $f_{\mathrm{fil}}$ parameter space more thoroughly. In reality, filament kinematics can vary. For example, observation of the Mon R2 region performed by \citet{monr2} show that filament accretion rates can range from 10$^{-4}$ - 10$^{-3}$M$_\odot$yr$^{-1}$ for filaments with line masses in the range 30-100M$_\odot$pc$^{-1}$. While filament velocities of $\approx 2$kms$^{-1}$ can lie in this range, it can also correspond to filaments with velocities $< 1$kms$^{-1}$. Furthermore, gradients along filament widths have been observed in filaments around the Serpens South cluster (\citealt{fernandez}, \citealt{dhabal}, \citealt{chen_fil}) implying even more complex kinematics along filaments. In this work, we did not explore the parameter space spanned by kinematics. However, because lower filament velocities result in lower kinetic energy injection into the central cluster, we can expect that less gas would be removed from the central cluster is the filament velocities were lower in our simulations. As well, probing lower filament velocities would lead to less gas mass being accreted onto the cluster. These lower velocity filaments would, however, have a chance to form stars that could be accreted onto the cluster at later times.

As well, there exists evidence of a commonality amongst filaments showing that they have characteristic widths of $\approx0.1$pc (\citealt{andre_01}). In observations, this value is derived as the full width half maximum (FWHM) of a filament density distribution (\citealt{arz2011}). Using the entire filament density distribution as the filament width has been shown to result in widths of 0.3pc (\citealt{SDC}) which we used in our simulations (see figure \ref{fig:fil_dens_ex} for an example). Though our simulations do not consider the FWHM of the filament density distribution as its width, we can still extrapolate how our simulations would behave if our filaments had a width of 0.1pc. Because our results show that the width of a filament with respect to the size of the cluster is important is determining how the cluster will react to accretion, we conclude that accretion from filaments with a width of 0.1pc will result in less gas mass becoming unbound from the cluster. Clusters accreting from filaments with a width of 0.1pc will also see an increase in the amount of potentially star forming dense gas over time. This would lead to an even better agreement with the sink particle prescription when calculating the total mass of the accreting cluster.

Our 30 simulations of gas accretion onto a central star cluster provide us with hints into the complexity of young cluster evolution inside GMCs. Coupling the results from this work with those from Paper I leads to a more comprehensive view of the sink particle prescription, and provides ways in which it can be improved in larger scale GMC simulations. We find that mergers of young gas rich subclusters do more to contribute to inaccuracies between our models and the sink particle prescription than accretion from ambient background gas or filaments. In the future, it is important to know how the two processes, mergers and gas accretion, affect one another directly. Our future work will probe this further and lead us to simulations of the full build up of massive clusters from H18, and similar simulations, in greater detail.

\section*{Acknowledgements}
The authors thank Claude Cournoyer-Cloutier, Michelle Nguyen, Marta Reina-Campos, Veronika Dornan, and Rachel Pillsworth for helpful discussions. The authors also thank the anonymous referee for careful reading and constructive comments. AS is supported by the Natural Sciences and Engineering Research Council of Canada. This research was enabled in part by support provided by Compute Ontario (https://www.computeontario.ca) and Compute Canada (http://www.computecanada.ca).

\section*{Data Availability}

The simulations described in this paper are available upon reasonable request to the corresponding author.



\bibliographystyle{mnras}
\bibliography{references} 

\begin{thebibliography}{}
\makeatletter
\relax
\def\mn@urlcharsother{\let\do\@makeother \do\$\do\&\do\#\do\^\do\_\do\%\do\~}
\def\mn@doi{\begingroup\mn@urlcharsother \@ifnextchar [ {\mn@doi@}
  {\mn@doi@[]}}
\def\mn@doi@[#1]#2{\def\@tempa{#1}\ifx\@tempa\@empty \href
  {http://dx.doi.org/#2} {doi:#2}\else \href {http://dx.doi.org/#2} {#1}\fi
  \endgroup}
\def\mn@eprint#1#2{\mn@eprint@#1:#2::\@nil}
\def\mn@eprint@arXiv#1{\href {http://arxiv.org/abs/#1} {{\tt arXiv:#1}}}
\def\mn@eprint@dblp#1{\href {http://dblp.uni-trier.de/rec/bibtex/#1.xml}
  {dblp:#1}}
\def\mn@eprint@#1:#2:#3:#4\@nil{\def\@tempa {#1}\def\@tempb {#2}\def\@tempc
  {#3}\ifx \@tempc \@empty \let \@tempc \@tempb \let \@tempb \@tempa \fi \ifx
  \@tempb \@empty \def\@tempb {arXiv}\fi \@ifundefined
  {mn@eprint@\@tempb}{\@tempb:\@tempc}{\expandafter \expandafter \csname
  mn@eprint@\@tempb\endcsname \expandafter{\@tempc}}}

\bibitem[\protect\citeauthoryear{{Aarseth}}{{Aarseth}}{1974}]{aarseth}
{Aarseth} S.~J.,  1974, \aap, \href
  {https://ui.adsabs.harvard.edu/abs/1974A&A....35..237A} {35, 237}

\bibitem[\protect\citeauthoryear{{Andr{\'e}}}{{Andr{\'e}}}{2017}]{andre_SF}
{Andr{\'e}} P.,  2017, \mn@doi [Comptes Rendus Geoscience]
  {10.1016/j.crte.2017.07.002}, \href
  {https://ui.adsabs.harvard.edu/abs/2017CRGeo.349..187A} {349, 187}

\bibitem[\protect\citeauthoryear{{Andr{\'e}} et~al.,}{{Andr{\'e}}
  et~al.}{2010}]{andre_2010}
{Andr{\'e}} P.,  et~al., 2010, \mn@doi [\aap] {10.1051/0004-6361/201014666},
  \href {https://ui.adsabs.harvard.edu/abs/2010A&A...518L.102A} {518, L102}

\bibitem[\protect\citeauthoryear{{Andr{\'e}}, {Di Francesco}, {Ward-Thompson},
  {Inutsuka}, {Pudritz}  \& {Pineda}}{{Andr{\'e}} et~al.}{2014}]{andre}
{Andr{\'e}} P.,  {Di Francesco} J.,  {Ward-Thompson} D.,  {Inutsuka} S.~I.,
  {Pudritz} R.~E.,   {Pineda} J.~E.,  2014, in {Beuther} H.,  {Klessen} R.~S.,
  {Dullemond} C.~P.,   {Henning} T.,  eds, Protostars and Planets VI. p.~27
  (\mn@eprint {arXiv} {1312.6232}),
  \mn@doi{10.2458/azu\_uapress\_9780816531240-ch002}

\bibitem[\protect\citeauthoryear{{Andr{\'e}}, {Palmeirim}  \&
  {Arzoumanian}}{{Andr{\'e}} et~al.}{2022}]{andre_01}
{Andr{\'e}} P.,  {Palmeirim} P.,   {Arzoumanian} D.,  2022, arXiv e-prints,
  \href {https://ui.adsabs.harvard.edu/abs/2022arXiv221004736A} {p.
  arXiv:2210.04736}

\bibitem[\protect\citeauthoryear{{Arzoumanian} et~al.,}{{Arzoumanian}
  et~al.}{2011}]{arz2011}
{Arzoumanian} D.,  et~al., 2011, \mn@doi [\aap] {10.1051/0004-6361/201116596},
  \href {https://ui.adsabs.harvard.edu/abs/2011A&A...529L...6A} {529, L6}

\bibitem[\protect\citeauthoryear{{Barnes} \& {Hut}}{{Barnes} \&
  {Hut}}{1986}]{tree}
{Barnes} J.,  {Hut} P.,  1986, \mn@doi [Nature Astronomy] {10.1038/324446a0},
  \href {https://ui.adsabs.harvard.edu/abs/1986Natur.324..446B} {324, 446}

\bibitem[\protect\citeauthoryear{{Bhadari}, {Dewangan}, {Ojha}, {Pirogov}  \&
  {Maity}}{{Bhadari} et~al.}{2022}]{bhadari}
{Bhadari} N.~K.,  {Dewangan} L.~K.,  {Ojha} D.~K.,  {Pirogov} L.~E.,   {Maity}
  A.~K.,  2022, \mn@doi [\apj] {10.3847/1538-4357/ac65e9}, \href
  {https://ui.adsabs.harvard.edu/abs/2022ApJ...930..169B} {930, 169}

\bibitem[\protect\citeauthoryear{{Bleuler} \& {Teyssier}}{{Bleuler} \&
  {Teyssier}}{2014}]{new_sink}
{Bleuler} A.,  {Teyssier} R.,  2014, \mn@doi [\mnras] {10.1093\mnras/stu2005},
  \href {https://ui.adsabs.harvard.edu/abs/2014MNRAS.445.4015B} {445, 4015}

\bibitem[\protect\citeauthoryear{{Bondi}}{{Bondi}}{1952}]{bondi}
{Bondi} H.,  1952, \mn@doi [\mnras] {10.1093/mnras/112.2.195}, \href
  {https://ui.adsabs.harvard.edu/abs/1952MNRAS.112..195B} {112, 195}

\bibitem[\protect\citeauthoryear{{Calura}, {D'Ercole}, {Vesperini}, {Vanzella}
  \& {Sollima}}{{Calura} et~al.}{2019}]{calura}
{Calura} F.,  {D'Ercole} A.,  {Vesperini} E.,  {Vanzella} E.,   {Sollima} A.,
  2019, \mn@doi [\mnras] {10.1093/mnras/stz2055}, \href
  {https://ui.adsabs.harvard.edu/abs/2019MNRAS.489.3269C} {489, 3269}

\bibitem[\protect\citeauthoryear{{Chen} et~al.,}{{Chen}
  et~al.}{2019}]{chen_fil}
{Chen} H.-R.~V.,  et~al., 2019, \mn@doi [\apj] {10.3847/1538-4357/ab0f3e},
  \href {https://ui.adsabs.harvard.edu/abs/2019ApJ...875...24C} {875, 24}

\bibitem[\protect\citeauthoryear{{Chevance}, {Krumholz}, {McLeod}, {Ostriker},
  {Rosolowsky}  \& {Sternberg}}{{Chevance} et~al.}{2022}]{chevance_2}
{Chevance} M.,  {Krumholz} M.~R.,  {McLeod} A.~F.,  {Ostriker} E.~C.,
  {Rosolowsky} E.~W.,   {Sternberg} A.,  2022, arXiv e-prints, \href
  {https://ui.adsabs.harvard.edu/abs/2022arXiv220309570C} {p. arXiv:2203.09570}

\bibitem[\protect\citeauthoryear{{Dhabal}, {Mundy}, {Rizzo}, {Storm}  \&
  {Teuben}}{{Dhabal} et~al.}{2018}]{dhabal}
{Dhabal} A.,  {Mundy} L.~G.,  {Rizzo} M.~J.,  {Storm} S.,   {Teuben} P.,  2018,
  \mn@doi [\apj] {10.3847/1538-4357/aaa76b}, \href
  {https://ui.adsabs.harvard.edu/abs/2018ApJ...853..169D} {853, 169}

\bibitem[\protect\citeauthoryear{{Dobbs}, {Bending}, {Pettitt}  \&
  {Bate}}{{Dobbs} et~al.}{2022}]{dobbs_1}
{Dobbs} C.~L.,  {Bending} T.~J.~R.,  {Pettitt} A.~R.,   {Bate} M.~R.,  2022,
  \mn@doi [\mnras] {10.1093/mnras/stab3036}, \href
  {https://ui.adsabs.harvard.edu/abs/2022MNRAS.509..954D} {509, 954}

\bibitem[\protect\citeauthoryear{{Evans} Neal~J. et~al.,}{{Evans}
  et~al.}{2009}]{evans09}
{Evans} Neal~J. I.,  et~al., 2009, \mn@doi [\apjs]
  {10.1088/0067-0049/181/2/321}, \href
  {https://ui.adsabs.harvard.edu/abs/2009ApJS..181..321E} {181, 321}

\bibitem[\protect\citeauthoryear{{Federrath}, {Banerjee}, {Clark}  \&
  {Klessen}}{{Federrath} et~al.}{2010}]{sink}
{Federrath} C.,  {Banerjee} R.,  {Clark} P.~C.,   {Klessen} R.~S.,  2010,
  \mn@doi [\apj] {10.1088/0004-637X/713/1/269}, \href
  {https://ui.adsabs.harvard.edu/abs/2010ApJ...713..269F} {713, 269}

\bibitem[\protect\citeauthoryear{{Fern{\'a}ndez-L{\'o}pez}
  et~al.,}{{Fern{\'a}ndez-L{\'o}pez} et~al.}{2014}]{fernandez}
{Fern{\'a}ndez-L{\'o}pez} M.,  et~al., 2014, \mn@doi [\apjl]
  {10.1088/2041-8205/790/2/L19}, \href
  {https://ui.adsabs.harvard.edu/abs/2014ApJ...790L..19F} {790, L19}

\bibitem[\protect\citeauthoryear{{Fischera} \& {Martin}}{{Fischera} \&
  {Martin}}{2012}]{fischera}
{Fischera} J.,  {Martin} P.~G.,  2012, \mn@doi [\aap]
  {10.1051/0004-6361/201218961}, \href
  {https://ui.adsabs.harvard.edu/abs/2012A&A...542A..77F} {542, A77}

\bibitem[\protect\citeauthoryear{{Fujii}, {Iwasawa}, {Funato}  \&
  {Makino}}{{Fujii} et~al.}{2007}]{bridge}
{Fujii} M.,  {Iwasawa} M.,  {Funato} Y.,   {Makino} J.,  2007, \mn@doi [PASJ]
  {10.1093/pasj/59.6.1095}, \href
  {https://ui.adsabs.harvard.edu/abs/2007PASJ...59.1095F} {59, 1095}

\bibitem[\protect\citeauthoryear{{Fukui} et~al.,}{{Fukui}
  et~al.}{2019}]{fukui_papillon}
{Fukui} Y.,  et~al., 2019, \mn@doi [\apj] {10.3847/1538-4357/ab4900}, \href
  {https://ui.adsabs.harvard.edu/abs/2019ApJ...886...14F} {886, 14}

\bibitem[\protect\citeauthoryear{{G{\'o}mez} \&
  {V{\'a}zquez-Semadeni}}{{G{\'o}mez} \&
  {V{\'a}zquez-Semadeni}}{2014}]{gomez_fil_vel}
{G{\'o}mez} G.~C.,  {V{\'a}zquez-Semadeni} E.,  2014, \mn@doi [\apj]
  {10.1088/0004-637X/791/2/124}, \href
  {https://ui.adsabs.harvard.edu/abs/2014ApJ...791..124G} {791, 124}

\bibitem[\protect\citeauthoryear{{Grudi{\'c}}, {Guszejnov}, {Offner}, {Rosen},
  {Raju}, {Faucher-Gigu{\`e}re}  \& {Hopkins}}{{Grudi{\'c}}
  et~al.}{2022}]{starforge2}
{Grudi{\'c}} M.~Y.,  {Guszejnov} D.,  {Offner} S. S.~R.,  {Rosen} A.~L.,
  {Raju} A.~N.,  {Faucher-Gigu{\`e}re} C.-A.,   {Hopkins} P.~F.,  2022, \mn@doi
  [\mnras] {10.1093/mnras/stac526}, \href
  {https://ui.adsabs.harvard.edu/abs/2022MNRAS.512..216G} {512, 216}

\bibitem[\protect\citeauthoryear{{Guszejnov}, {Markey}, {Offner}, {Grudi{\'c}},
  {Faucher-Gigu{\`e}re}, {Rosen}  \& {Hopkins}}{{Guszejnov}
  et~al.}{2022}]{starforge1}
{Guszejnov} D.,  {Markey} C.,  {Offner} S. S.~R.,  {Grudi{\'c}} M.~Y.,
  {Faucher-Gigu{\`e}re} C.-A.,  {Rosen} A.~L.,   {Hopkins} P.~F.,  2022, arXiv
  e-prints, \href {https://ui.adsabs.harvard.edu/abs/2022arXiv220101781G} {p.
  arXiv:2201.01781}

\bibitem[\protect\citeauthoryear{{Heiderman}, {Evans}, {Allen}, {Huard}  \&
  {Heyer}}{{Heiderman} et~al.}{2010}]{heid10}
{Heiderman} A.,  {Evans} Neal~J. I.,  {Allen} L.~E.,  {Huard} T.,   {Heyer} M.,
   2010, \mn@doi [\apj] {10.1088/0004-637X/723/2/1019}, \href
  {https://ui.adsabs.harvard.edu/abs/2010ApJ...723.1019H} {723, 1019}

\bibitem[\protect\citeauthoryear{{Howard}, {Pudritz}  \& {Harris}}{{Howard}
  et~al.}{2018}]{Howard2018}
{Howard} C.~S.,  {Pudritz} R.~E.,   {Harris} W.~E.,  2018, \mn@doi [Nature
  Astronomy] {10.1038/s41550-018-0506-0}, \href
  {https://ui.adsabs.harvard.edu/abs/2018NatAs...2..725H} {2, 725}

\bibitem[\protect\citeauthoryear{{Inutsuka} \& {Miyama}}{{Inutsuka} \&
  {Miyama}}{1992}]{inutsuka_2}
{Inutsuka} S.-I.,  {Miyama} S.~M.,  1992, \mn@doi [\apj] {10.1086/171162},
  \href {https://ui.adsabs.harvard.edu/abs/1992ApJ...388..392I} {388, 392}

\bibitem[\protect\citeauthoryear{{Inutsuka} \& {Miyama}}{{Inutsuka} \&
  {Miyama}}{1997}]{inutsuka_1}
{Inutsuka} S.-i.,  {Miyama} S.~M.,  1997, \mn@doi [\apj] {10.1086/303982},
  \href {https://ui.adsabs.harvard.edu/abs/1997ApJ...480..681I} {480, 681}

\bibitem[\protect\citeauthoryear{{Kaaz}, {Antoni}  \& {Ramirez-Ruiz}}{{Kaaz}
  et~al.}{2019}]{kaaz}
{Kaaz} N.,  {Antoni} A.,   {Ramirez-Ruiz} E.,  2019, \mn@doi [\apj]
  {10.3847/1538-4357/ab158b}, \href
  {https://ui.adsabs.harvard.edu/abs/2019ApJ...876..142K} {876, 142}

\bibitem[\protect\citeauthoryear{{Karam} \& {Sills}}{{Karam} \&
  {Sills}}{2022}]{karam}
{Karam} J.,  {Sills} A.,  2022, \mn@doi [\mnras] {10.1093/mnras/stac1298},
  \href {https://ui.adsabs.harvard.edu/abs/2022MNRAS.513.6095K} {513, 6095}

\bibitem[\protect\citeauthoryear{{Kirk}, {Myers}, {Bourke}, {Gutermuth},
  {Hedden}  \& {Wilson}}{{Kirk} et~al.}{2013}]{kirk}
{Kirk} H.,  {Myers} P.~C.,  {Bourke} T.~L.,  {Gutermuth} R.~A.,  {Hedden} A.,
  {Wilson} G.~W.,  2013, \mn@doi [\apj] {10.1088/0004-637X/766/2/115}, \href
  {https://ui.adsabs.harvard.edu/abs/2013ApJ...766..115K} {766, 115}

\bibitem[\protect\citeauthoryear{{K{\"o}nyves} et~al.,}{{K{\"o}nyves}
  et~al.}{2010}]{koy}
{K{\"o}nyves} V.,  et~al., 2010, \mn@doi [\aap] {10.1051/0004-6361/201014689},
  \href {https://ui.adsabs.harvard.edu/abs/2010A&A...518L.106K} {518, L106}

\bibitem[\protect\citeauthoryear{{Kroupa}}{{Kroupa}}{2001}]{kroupa2001}
{Kroupa} P.,  2001, \mn@doi [\mnras] {10.1046/j.1365-8711.2001.04022.x}, \href
  {https://ui.adsabs.harvard.edu/abs/2001MNRAS.322..231K} {322, 231}

\bibitem[\protect\citeauthoryear{{Krumholz}, {McKee}  \& {Klein}}{{Krumholz}
  et~al.}{2004}]{krumholz_sink}
{Krumholz} M.~R.,  {McKee} C.~F.,   {Klein} R.~I.,  2004, \mn@doi [\apj]
  {10.1086/421935}, \href
  {https://ui.adsabs.harvard.edu/abs/2004ApJ...611..399K} {611, 399}

\bibitem[\protect\citeauthoryear{{Kumar}, {Arzoumanian}, {Men'shchikov},
  {Palmeirim}, {Matsumura}  \& {Inutsuka}}{{Kumar} et~al.}{2022}]{kumar}
{Kumar} M.~S.~N.,  {Arzoumanian} D.,  {Men'shchikov} A.,  {Palmeirim} P.,
  {Matsumura} M.,   {Inutsuka} S.,  2022, \mn@doi [\aap]
  {10.1051/0004-6361/202140363}, \href
  {https://ui.adsabs.harvard.edu/abs/2022A&A...658A.114K} {658, A114}

\bibitem[\protect\citeauthoryear{{Lada} \& {Lada}}{{Lada} \&
  {Lada}}{2003}]{ladalada}
{Lada} C.~J.,  {Lada} E.~A.,  2003, \mn@doi [Annual Reviews of \aap]
  {10.1146/annurev.astro.41.011802.094844}, \href
  {https://ui.adsabs.harvard.edu/abs/2003ARA&A..41...57L} {41, 57}

\bibitem[\protect\citeauthoryear{{Lada}, {Lombardi}  \& {Alves}}{{Lada}
  et~al.}{2010}]{lada2010}
{Lada} C.~J.,  {Lombardi} M.,   {Alves} J.~F.,  2010, \mn@doi [\apj]
  {10.1088/0004-637X/724/1/687}, \href
  {https://ui.adsabs.harvard.edu/abs/2010ApJ...724..687L} {724, 687}

\bibitem[\protect\citeauthoryear{{Lada}, {Forbrich}, {Lombardi}  \&
  {Alves}}{{Lada} et~al.}{2012}]{lada2012}
{Lada} C.~J.,  {Forbrich} J.,  {Lombardi} M.,   {Alves} J.~F.,  2012, \mn@doi
  [\apj] {10.1088/0004-637X/745/2/190}, \href
  {https://ui.adsabs.harvard.edu/abs/2012ApJ...745..190L} {745, 190}

\bibitem[\protect\citeauthoryear{{Lee}, {Looney}, {Schnee}  \& {Li}}{{Lee}
  et~al.}{2013}]{lee}
{Lee} K.,  {Looney} L.~W.,  {Schnee} S.,   {Li} Z.-Y.,  2013, \mn@doi [\apj]
  {10.1088/0004-637X/772/2/100}, \href
  {https://ui.adsabs.harvard.edu/abs/2013ApJ...772..100L} {772, 100}

\bibitem[\protect\citeauthoryear{{Li}, {Vogelsberger}, {Marinacci}  \&
  {Gnedin}}{{Li} et~al.}{2019}]{li2019}
{Li} H.,  {Vogelsberger} M.,  {Marinacci} F.,   {Gnedin} O.~Y.,  2019, \mn@doi
  [\mnras] {10.1093\mnras/stz1271}, \href
  {https://ui.adsabs.harvard.edu/abs/2019MNRAS.487..364L} {487, 364}

\bibitem[\protect\citeauthoryear{{Makino} \& {Aarseth}}{{Makino} \&
  {Aarseth}}{1992}]{hermite}
{Makino} J.,  {Aarseth} S.~J.,  1992, PASJ, \href
  {https://ui.adsabs.harvard.edu/abs/1992PASJ...44..141M} {44, 141}

\bibitem[\protect\citeauthoryear{{Men'shchikov} et~al.,}{{Men'shchikov}
  et~al.}{2010}]{mens}
{Men'shchikov} A.,  et~al., 2010, \mn@doi [\aap] {10.1051/0004-6361/201014668},
  \href {https://ui.adsabs.harvard.edu/abs/2010A&A...518L.103M} {518, L103}

\bibitem[\protect\citeauthoryear{{Naiman}, {Ramirez-Ruiz}  \& {Lin}}{{Naiman}
  et~al.}{2011}]{naiman}
{Naiman} J.~P.,  {Ramirez-Ruiz} E.,   {Lin} D.~N.~C.,  2011, \mn@doi [\apj]
  {10.1088/0004-637X/735/1/25}, \href
  {https://ui.adsabs.harvard.edu/abs/2011ApJ...735...25N} {735, 25}

\bibitem[\protect\citeauthoryear{{Ostriker}}{{Ostriker}}{1964}]{ostriker}
{Ostriker} J.,  1964, \mn@doi [\apj] {10.1086/148005}, \href
  {https://ui.adsabs.harvard.edu/abs/1964ApJ...140.1056O} {140, 1056}

\bibitem[\protect\citeauthoryear{{Pelupessy} \& {Portegies Zwart}}{{Pelupessy}
  \& {Portegies Zwart}}{2012}]{pelupessy2013}
{Pelupessy} F.~I.,  {Portegies Zwart} S.,  2012, \mn@doi [\mnras]
  {10.1111/j.1365-2966.2011.20137.x}, \href
  {https://ui.adsabs.harvard.edu/abs/2012MNRAS.420.1503P} {420, 1503}

\bibitem[\protect\citeauthoryear{{Peretto} et~al.,}{{Peretto}
  et~al.}{2014}]{SDC}
{Peretto} N.,  et~al., 2014, \mn@doi [\aap] {10.1051/0004-6361/201322172},
  \href {https://ui.adsabs.harvard.edu/abs/2014A&A...561A..83P} {561, A83}

\bibitem[\protect\citeauthoryear{{Plummer}}{{Plummer}}{1911}]{plummer1911}
{Plummer} H.~C.,  1911, \mn@doi [\mnras] {10.1093\mnras/71.5.460}, \href
  {https://ui.adsabs.harvard.edu/abs/1911MNRAS..71..460P} {71, 460}

\bibitem[\protect\citeauthoryear{{Portegies Zwart} et~al.,}{{Portegies Zwart}
  et~al.}{2009}]{zwart2009}
{Portegies Zwart} S.,  et~al., 2009, \mn@doi [New Astronomy]
  {10.1016/j.newast.2008.10.006}, \href
  {https://ui.adsabs.harvard.edu/abs/2009NewA...14..369P} {14, 369}

\bibitem[\protect\citeauthoryear{{Portegies Zwart}, {McMillan}  \&
  {Gieles}}{{Portegies Zwart} et~al.}{2010}]{zwart}
{Portegies Zwart} S.~F.,  {McMillan} S. L.~W.,   {Gieles} M.,  2010, \mn@doi
  [Annual Reviews of \aap] {10.1146/annurev-astro-081309-130834}, \href
  {https://ui.adsabs.harvard.edu/abs/2010ARA&A..48..431P} {48, 431}

\bibitem[\protect\citeauthoryear{{Rieder}, {Dobbs}, {Bending}, {Liow}  \&
  {Wurster}}{{Rieder} et~al.}{2022}]{ekster}
{Rieder} S.,  {Dobbs} C.,  {Bending} T.,  {Liow} K.~Y.,   {Wurster} J.,  2022,
  \mn@doi [\mnras] {10.1093/mnras/stab3425}, \href
  {https://ui.adsabs.harvard.edu/abs/2022MNRAS.509.6155R} {509, 6155}

\bibitem[\protect\citeauthoryear{{Rosolowsky} et~al.,}{{Rosolowsky}
  et~al.}{2021}]{rosolowsky}
{Rosolowsky} E.,  et~al., 2021, \mn@doi [\mnras] {10.1093/mnras/stab085}, \href
  {https://ui.adsabs.harvard.edu/abs/2021MNRAS.502.1218R} {502, 1218}

\bibitem[\protect\citeauthoryear{{Shima}, {Matsuda}, {Takeda}  \&
  {Sawada}}{{Shima} et~al.}{1985}]{shima}
{Shima} E.,  {Matsuda} T.,  {Takeda} H.,   {Sawada} K.,  1985, \mn@doi [\mnras]
  {10.1093/mnras/217.2.367}, \href
  {https://ui.adsabs.harvard.edu/abs/1985MNRAS.217..367S} {217, 367}

\bibitem[\protect\citeauthoryear{{Springel}}{{Springel}}{2005}]{gadget2}
{Springel} V.,  2005, \mn@doi [\mnras] {10.1111/j.1365-2966.2005.09655.x},
  \href {https://ui.adsabs.harvard.edu/abs/2005MNRAS.364.1105S} {364, 1105}

\bibitem[\protect\citeauthoryear{{Trevi{\~n}o-Morales}
  et~al.,}{{Trevi{\~n}o-Morales} et~al.}{2019}]{monr2}
{Trevi{\~n}o-Morales} S.~P.,  et~al., 2019, \mn@doi [\aap]
  {10.1051/0004-6361/201935260}, \href
  {https://ui.adsabs.harvard.edu/abs/2019A&A...629A..81T} {629, A81}

\bibitem[\protect\citeauthoryear{{Wang} et~al.,}{{Wang} et~al.}{2022}]{SDC_2}
{Wang} J.-W.,  et~al., 2022, \mn@doi [\apj] {10.3847/1538-4357/ac6872}, \href
  {https://ui.adsabs.harvard.edu/abs/2022ApJ...931..115W} {931, 115}

\bibitem[\protect\citeauthoryear{{White}}{{White}}{1994}]{white_glass}
{White} S. D.~M.,  1994, arXiv e-prints, \href
  {https://ui.adsabs.harvard.edu/abs/1994astro.ph.10043W} {pp
  astro--ph/9410043}

\bibitem[\protect\citeauthoryear{{Williams}, {Peretto}, {Avison},
  {Duarte-Cabral}  \& {Fuller}}{{Williams} et~al.}{2018}]{jvla}
{Williams} G.~M.,  {Peretto} N.,  {Avison} A.,  {Duarte-Cabral} A.,   {Fuller}
  G.~A.,  2018, \mn@doi [\aap] {10.1051/0004-6361/201731587}, \href
  {https://ui.adsabs.harvard.edu/abs/2018A&A...613A..11W} {613, A11}

\bibitem[\protect\citeauthoryear{{Wong} et~al.,}{{Wong} et~al.}{2022}]{30_dor}
{Wong} T.,  et~al., 2022, arXiv e-prints, \href
  {https://ui.adsabs.harvard.edu/abs/2022arXiv220606528W} {p. arXiv:2206.06528}

\bibitem[\protect\citeauthoryear{{Yuan} et~al.,}{{Yuan} et~al.}{2018}]{yuan}
{Yuan} J.,  et~al., 2018, \mn@doi [\apj] {10.3847/1538-4357/aa9d40}, \href
  {https://ui.adsabs.harvard.edu/abs/2018ApJ...852...12Y} {852, 12}

\makeatother
\end{thebibliography}




\appendix


\bsp	
\label{lastpage}
\end{document}